\documentclass[]{pasj01}

\Received{2022/06/13}
\Accepted{2022/09/04}
 
\begin{document} 

\title{ 
The soft X-ray background with Suzaku: I. Milky Way halo
}

 \author{Masaki \textsc{Ueda}\altaffilmark{1}, Hayato \textsc{Sugiyama}\altaffilmark{1}, Shogo B. \textsc{Kobayashi}\altaffilmark{1},Kotaro \textsc{Fukushima}\altaffilmark{1}, Noriko Y. \textsc{Yamasaki}\altaffilmark{2}, Kosuke \textsc{Sato} \altaffilmark{3}, and  Kyoko \textsc{Matsushita}\altaffilmark{1,}$^{*}$}
\altaffiltext{1}{Department of Physics, Tokyo University of Science, 1-3 Kagurazaka, Shinjuku-ku, Tokyo 162-8601 Japan}
\altaffiltext{2}{ Institute of Space and Astronautical Science, Japan Aerospace Exploration Agency, 3-1-1 Yoshinodai, Chuo-ku, Sagamihara, Kanagawa 252-5210, Japan}
\altaffiltext{3}{Graduate School of Science and Engineering, Saitama University, 255 Shimo-Okubo, Sakura-ku, Saitama, Saitama 338-8570, Japan} 
\email{matusita@rs.tus.ac.jp}



\maketitle

\begin{abstract}
We present measurements of the soft X-ray background emission for 130 Suzaku
observations at $75^\circ<l < 285^\circ$ and $|b|>15^\circ$ obtained from 2005 to 2015, covering nearly one solar cycle.
In addition to the standard soft X-ray background model consisting of the local hot bubble and the Milky Way Halo (MWH), we include a hot collisional-ionization-equilibrium component with a temperature of  $\sim 0.8$ keV to reproduce spectra of a significant fraction of the lines of sight.
Then, the scatter in the relation between the emission measure vs. temperature of the MWH component is reduced.
Here, we exclude time ranges with high count rates to minimize the effect of the solar wind charge exchange (SWCX). However, the spectra of almost the same lines of sight are inconsistent. The heliospheric SWCX emissions likely contaminate and gives a bias in measurements of temperature and the emission measure of the MWH.
Excluding the data around the solar maximum and using the data taken before the end of 2009, 
at $|b|>35^\circ$ and $105^\circ<l<255^\circ$, 
the temperature (0.22 keV) and emission measure ($2\times 10^{-3}~\rm{cm^{-6}pc}$) of the MWH are fairly uniform.
The increase of the emission measure toward the lower Galactic latitude at $|b|<35^\circ$ 
indicates a presence of a disk-like morphology component.
A composite model which consists of disk-like and spherical-morphology components also reproduces the observed emission measure distribution of MWH.
In this case, the hydrostatic mass at a few tens of kpc from the Galactic center  agrees with the gravitational mass of the Milky Way.
The plasma with the virial temperature likely fills the Milky Way halo in nearly hydrostatic equilibrium.
Assuming the gas metallicity of 0.3 solar,  the upper limit of the gas mass of the spherical component out to 250 kpc, or the virial radius, is $\sim$  a few $\times 10^{10}~ M_\odot$.
\end{abstract}

\section{Introduction}

Galaxies are thought to be surrounded by an extended gaseous halo, the circumgalactic medium (CGM), which is expected to contain a significant fraction of the baryonic mass (e.g. \cite{CGM2017} and references therein).
Under the cold dark matter cosmology, 
galaxies grow by accretion of matter from their surroundings. 
The accreting gas is expected to be heated to the virial temperature 
by shock waves generated by mass accretion (e.g., \cite{Rees1977}) and fills the halo of galaxies in nearly hydrostatic equilibrium.
In addition, supernovae (SNe) heat the interstellar medium and sometimes cause an outflow toward intergalactic space.
Therefore, the observations of hot gas in galaxies are essential to study the structure formation of the Universe and feedback from galaxies to intergalactic space.
Suzaku, Japan's X-ray astronomy satellite, is characterized by its low background, enabling us to detect low-surface brightness X-ray emissions from hot gas in the Milky Way.

The soft X-ray background detected below $\sim $ 1 keV is thought to mainly
come from the hot gas in and around the Milky Way (e.g., \cite{Snowden1998}, \cite{Kuntz2000}).
The X-ray spectrum of the soft X-ray background is usually modeled with a sum of collisional-ionization-equilibrium (CIE) plasmas of 0.1 keV and 0.2--0.3 keV.
A cavity surrounding the sun filled with CIE plasma with a temperature of 0.1 keV is called the Local Hot Bubble (LHB).
The Milky Way Halo (MWH) component,  hot gas with a temperature of 0.2--0.3 keV, probably comes from a more extended, diffuse plasma.
The emission measure of the MWH component varies by over an order of magnitude, even toward the Galactic anticenter (\cite{Henley13}, \cite{Nakashima18}).
The detection of  absorption lines of O\,\emissiontype{VII} and O\,\emissiontype{VIII}
in the X-ray spectra of active galactic nuclei (e.g. \cite{Nicastro2002}, \cite{Fang2003}, \cite{Hagihara2010}) confirmed the presence of 
0.2--0.3 keV plasma around the Milky Way.
In addition,  \citet{Yoshino09}, \citet{Henley13}, \citet{Sekiya14a}, \citet{Nakashima18}, \citet{Gupta21},  \citet{Gupta2022}   reported "excess" emissions which can be modeled  by a plasma with supersolar Ne/Fe ratio or a higher temperature component (0.5-0.9 keV) from several lines of sight.

Heliospheric and geocoronal solar wind charge exchange (SWCX) emissions sometimes
contaminate the soft X-ray emissions from outside the solar system.
The SWCX emission is caused by interaction between the solar wind ions and neutral
atoms (e.g. \cite{Cravens01},  \cite{Kout07}).
In some cases, highly time-variable strong emission lines of C, O, Ne, and Mg 
possibly caused by geocoronal SWCX were detected with XMM-Newton (e.g. \cite{Carter10})  and Suzaku (e.g. \cite{Fujimoto07}, \cite{Ishi19}). 
These emissions can be filtered using light curves in the soft energy band
and ion flux measurements of the solar wind.
The time variation of the heliospheric SWCX should be much slower, and its intensity is expected to relate to the solar cycle.
\citet{Yoshitake13} studied O\,\emissiontype{VII} line intensity toward the Lockman hole observed with Suzaku from 2006 and 2011.
After screening the geocoronal SWCX emissions using the soft-band light curve and proton flux of the solar wind, they found an O\,\emissiontype{VII} line enhancement possibly related to the heliospheric SWCX.
\citet{Qu22} measured the O\,\emissiontype{VII} and O\,\emissiontype{VIII} line strengths using the XMM data from 2000 to 2010 and studied the correlation with the solar activity. They concluded that the heliospheric SWCX emissions significantly (30--50 \%) contaminate these O lines in average.
In addition to the SWCX components, Suzaku data was contaminated with the strong O\,\emissiontype{I} K$\alpha$ line scattered by the earth's atmosphere around the solar maximum \citep{Sekiya14OI}.

In this study, we present data analysis of 130 observations with Suzaku at $75^\circ < l < 285^\circ$ and at $|b| > 15^\circ$ excluding the Galactic center and the Galactic plane to study the distribution of the hot gas which fills the Milky Way halo.
Here, $l$ is the Galactic longitude, and $b$ is the Galactic latitude.
These data were obtained from 2005 to 2015, covering nearly one solar cycle.
Some regions are observed multiple times and therefore suitable for studying the contaminations by SWCX.
In Section 2, we describe the observations and data reduction.
The spectral analysis and results are presented in Section 3.
We present the results of spectral fittings with an extra hot CIE component with a temperature of 0.8 keV.
This paper mainly shows the temperatures and emission measures of the MWH component.
Those of the 0.8 keV component will be presented in another paper (Sugiyama et al. in preparation).
We discuss and summarize the results in Sections 4 and 5, respectively.
This paper uses the solar abundance table by \citet{Lodders03}.
Errors are reported at the 68\% confidence level unless otherwise stated.

\section{Observations and data reduction}

We analyzed archival Suzaku/XIS data of 130 observations toward $75^\circ<l<285^\circ$ and $|b|>15^\circ$. The observation log is shown in Appendix.
Here, we excluded regions around extended objects, such as clusters of galaxies and supernova remnants.  The observations around very bright point sources are also excluded.
The sample includes ten data toward Lockman Hole from 2005 to 2014,  six data toward the North ecliptic pole (NEP) obtained in 2005, 2006, and 2009, and four data toward the South ecliptic pole (SEP) observed in 2009.
The pointing position of the Lockman Hole observation in 2005 is offset by 0.5$^\circ$ from the other Lockman Hole observations. Those of NEP in 2005 and 2006 are nearly identical, but the other four are offset by $\sim$ 1.2$^\circ$.
These observations with almost the same sightlines are suitable to constrain contaminating emissions caused by solar activity.

The XIS has four CCD sensors:
XIS0, 2, and 3 contain front-illuminated (FI) CCDs and
XIS1 has a back-illuminated (BI) CCD \citep{Koyama2007}. 
The Suzaku/XIS detectors are sensitive in the 0.4--10 keV energy band with 
the field of view (FOV) of $\sim 18'\times 18'$.
We analyzed the data of the four CCDs taken before the XIS2 loss
in November 2006. After the loss,  we analyzed the data of the three remaining CCDs.
We used data with normal clocking mode with no window option.
Data obtained with 3$\times$3 and 5$\times$5 editing modes were merged, and the standard filtering procedures (Earth elevation $>$ 10$^\circ$, cutoff rigidity $>$ 6 GeV/c) were applied. Additional flickering pixels\footnote{http://www.astro.isas.ac.jp/suzaku/analysis/xis/nxb\_new2/} were removed.

\begin{figure}
 \includegraphics[width=1.0\columnwidth]{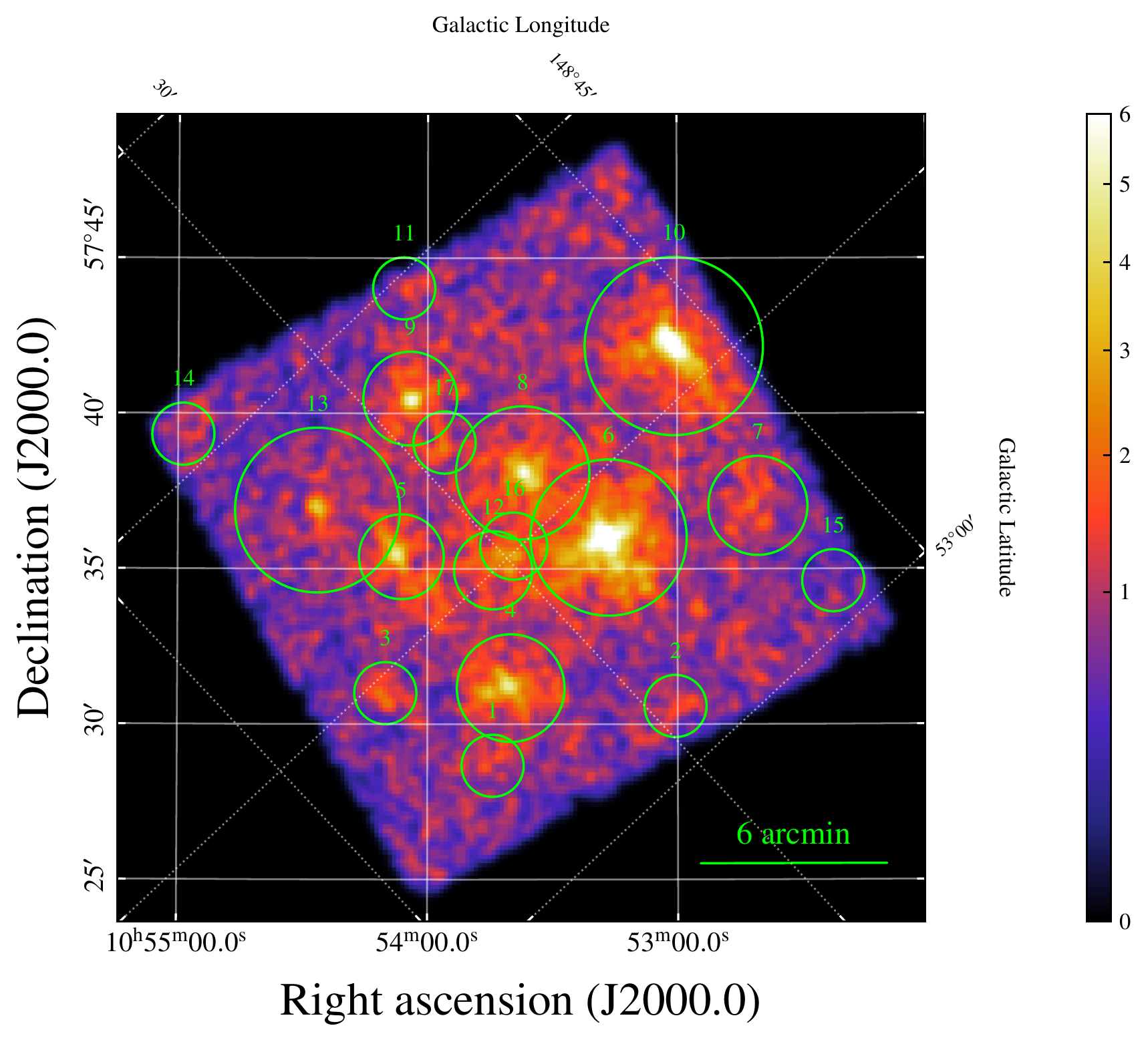}
  \caption{The XIS1 image (0.5--5.0 keV) of Lockman Hole (OBSID:100046010) observed in 2005. The circular regions with numbers are excluded from the spectral analysis.}\label{fig:lockmanimage}
\end{figure}

To remove point sources in the XIS FOVs, we created images in the 0.5--2.0 keV and 2.0--5.0 keV energy bands and applied the 
 \textit{wavdetect} tool in the CIAO package.
From the subsequent analysis, we excised circular regions around the point source candidates with a significance of $>3 \sigma$.
For sources with a flux lower than $5.0\times 10^{-14}~\rm{erg~s^{-1}cm^{-2}}$ in the energy band of 0.5--5.0 keV, we excised the circular region centered on each source with a radius of 1$'$.
For brighter sources, this radius is not enough to exclude scattered photons by the X-ray telescopes of Suzaku.
For example, two stars emit strong Fe-L lines,  numbered 8 and 10 in figure \ref{fig:lockmanimage}, in the FOV of the Lockman hole observation in 2005. 
Scattered photons from these stars strongly contaminate the spectra when the exclusion radius is 1$'$.
Therefore, for each brighter source,  we determined the exclusion radius based on its flux in the 0.5--7.0 keV band.
For example, the exclusion radii for the flux of $1.0\times 10^{-13}~\rm{erg~s^{-1}cm^{-2}}$ and $5.0\times 10^{-12}~\rm{erg~s^{-1}cm^{-2}}$  are 1$'$.5 and 4$'$.5, respectively.
For sources with a power-law spectrum of a photon index of 1.7, 
these radii correspond to the flux level of scattered photons from the point source is half of the background.
Further increasing the exclusion radii did not change the results.

To decrease the contamination caused  by the SWCX and other background fluctuations, we created light curves  
and excluded the time ranges when the count rate exceeds 3 $\sigma$.
To increase the available exposure time, we have not adopted any further filtering using the proton flux of the solar wind to reduce the effect of the SWCX 
or using elevation angle from the sunlit Earth's limb for suppressing contamination by the O\,\emissiontype{I} emission from the Earth's atmosphere.
The screening criteria used in this work are similar to those used to study low-surface brightness targets and, therefore, suitable to study the effects of the systematic uncertainties in other data obtained with Suzaku.

\section{Spectral analysis and results}

We extracted a spectrum over the FOV of each XIS detector of each observation.
To create ancillary response files (ARFs), we used the {\it xissimarfgen} ftools task \citep{Ishisaki07} assuming uniform emission from a circular region with a radius of 20$'$.
The effects of contaminants and the optical blocking filter of XISs were included in the ARFs.
We used the {\it xisrmfgen} ftools task to create the redistribution matrix files (RMFs).
The instrumental background, or non-X-ray background (NXB), was estimated from the night-Earth database
using the {\it xisnxbgen} ftools task \citep{Tawa08}.
We used XSPEC version 12.10.1b to model the NXB-subtracted spectra.
Unless otherwise stated, we
 used energy ranges of  0.4--7.0 keV  for the BI spectra, and 0.5--7.0 keV 
  for the FI spectra.
We rebinned each spectrum to a minimum of one count per bin and employed the extended C-statistic \citep{Cash79}.
We used APEC (Astrophysical Plasma Emission Code, \cite{Smith01}, \cite{Foster12}) with AtomDB version 3.0.9 to model a CIE plasma.

\subsection{Stacked Spectra of the 130 observations}

To study possible emission components of the soft X-ray background, 
we created stacked XIS-FI (XIS0, 2, 3) and XIS-BI (XIS1) spectra of the 130 observations.
We first tried a spectral fitting with a model which
consists of four components:
 the cosmic X-ray background (CXB),  two CIE components ({\it apec} model in XSPEC) to model  the LHB (and SWCX)  and MWH,
 and the O\,\emissiontype{I} K$\alpha$ line for the scattered photon from the sunlit atmosphere of the Earth.
Hereafter, we denote this model as Model-s.
 We adopted a power-law model with a photon index of 1.4 to model the CXB component.
The emission from the heliospheric SWCX and LHB are empirically modeled with an unabsorbed CIE component of  $kT\sim 0.1$  keV with the solar metallicity (e.g. \cite{Fujimoto07}, \cite{Yoshino09}, \cite{Henley13}, \cite{Nakashima18}).
We also fixed the temperature of the LHB component at 0.1 keV and 
the abundances of the LHB and MWH components at 1 solar.
 The CXB and the MWH components are subject to photoelectric absorption due to cold gas, and we modeled 
 this absorption using the {\it phabs} model in the XSPEC spectral fitting tool.    
 A Gaussian at fixed central energy of 0.525 keV was used to model the O\,\emissiontype{I} line.
 The normalization of each component was allowed to vary.
 Hereafter, we call the temperature and emission measure ($\int n_{\rm e}n_{\rm H} ds$)  of the MWH component as $kT_{\rm halo}$ and $\mathrm{EM}_{\rm halo}$, respectively.
 Here, $s$ is the distance along the line of sight.
    
\begin{figure}
  \includegraphics[width=0.9\columnwidth]{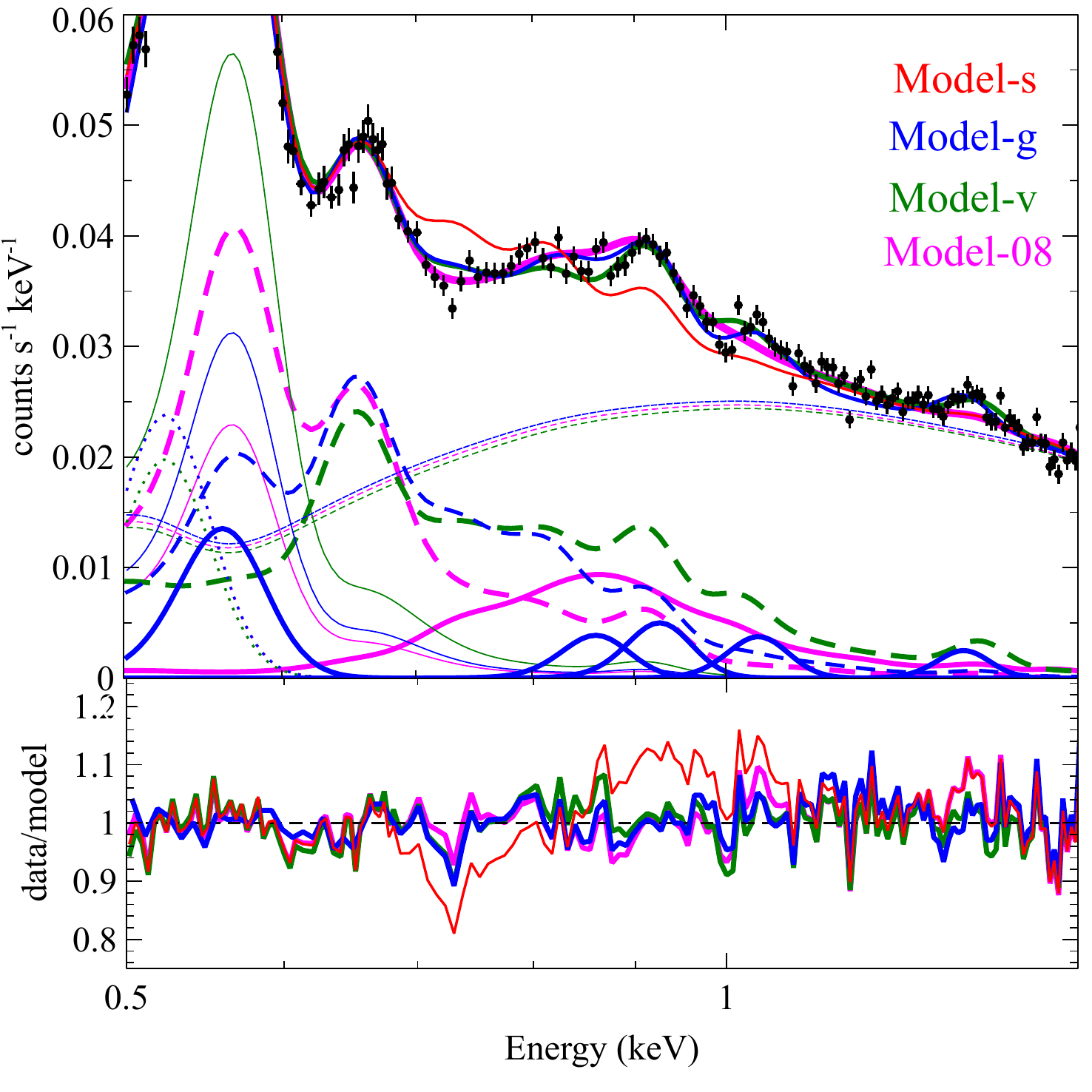} 
  \caption{ The stacked XIS1 spectra (top) and data-to-model ratio (bottom) fitted with Model-s(red), Model-g (blue), Mosel-v(green), and Model-08 (magenta). 
The contributions of MWH (thick dashed lines), LHB (thin solid lines), CXB (thin dashed lines), and O\,\emissiontype{I} (thin dotted lines) components  for Model-g, Model-v, and Model-08  are shown. The thick magenta solid line corresponds to the 0.8 keV component for Model-08, and the thick blue solid lines are additional Gaussians for Model-g.  }\label{fig:stacked}
\end{figure}

 We fitted the stacked BI and FI spectra simultaneously with Model-s.
 Here, we used energy ranges
 of 0.5--7.0 keV and 0.6--7.0 keV for the BI and FI spectra, respectively,
 since stacking may cause some problems in the lower energy band due to stacking spectra with 
 different column densities of the contamination and the Galactic absorption.
 Since there is a slight discrepancy between the stacked FI and BI spectra in the lower energy band, 
 we allowed the $N_{\rm H}$ for the two spectra to vary separately. 
The result of the spectral fitting with Model-s is shown in table \ref{tab:stacked} and the stacked BI spectrum and the best-fit model are shown in figure \ref{fig:stacked}.
This fit cannot reproduce the spectra with a C-statistics/d.o.f of 6843/3524 and there remain clear residual structures at 0.7--1 keV.

\begin{table*}
\tbl{Fitting results of the stacked spectra }{%
\begin{tabular}{lllll}
\hline
               & Model-s & Model-g & Model-v & Model-08\\\hline
$\mathrm{EM}_{\rm LHB}$\footnotemark[$*$]  & 2.3$\times 10^{-2}$&  $(1.3\pm 0.2)\times 10^{-2}$ & $(2.7\pm 0.1)\times 10^{-2}$ & $(1.1\pm 0.1)\times 10^{-2}$ \\
$kT_{\rm halo}$ (keV) & 0.24 & $0.219^{+0.001}_{-0.001}$ 
& $0.333\pm 0.007$ &  $0.175\pm 0.003$ \\
 $\mathrm{EM}_{\rm halo}$\footnotemark[$*$] & 2.9$\times 10^{-3}$ & $(3.4\pm 0.1)\times 10^{-3}$ 
 & $ (3.6\pm 0.5)\times 10^{-3}$   &$ (5.7\pm 0.3)\times 10^{-3}$ \\ 
MWH: O  (solar)    & 1.0 (fix)  & 1.0 (fix) & $0.91^{+0.19}_{-0.14}$  & 1.0 (fix)  \\
MWH: Ne (solar)    & 1.0 (fix)  & 1.0 (fix) & $0.81^{+0.15}_{-0.11}$ & 1.0 (fix)  \\
MWH: Mg  (solar)  & 1.0 (fix)  & 1.0 (fix)  & $0.78^{+0.13}_{-0.10}$  &  1.0 (fix) \\
MWH: Fe (solar)    & 1.0 (fix)  & 1.0 (fix) &   $0.21^{+0.04}_{-0.03}$  & 1.0 (fix) \\
O\,\emissiontype{I} Normalization\footnotemark[$\dagger$] & 1.7 & $1.72\pm 0.05 $   &   $1.50\pm 0.07$ &$1.80 \pm 0.05$   \\
$N_{\rm H}$ (10$^{20}\rm{cm}^{-2}$) for FI &0.0 & 0.0 ($<0.1)$ & 0.0 ($<0.1)$  &  0.0 ($<0.1$)\\
$N_{\rm H}$ (10$^{20}\rm{cm}^{-2}$) for BI &2.6 & 2.4$\pm 0.3$ & 3.3$\pm 0.3$ &2.6$\pm 0.3$   \\
CXB norm\footnotemark[$\ddagger$] & 8.7 & $8.55\pm 0.03$ & $8.44\pm 0.03$  & $8.46\pm 0.03$\\
$kT_{\rm 0.8 keV}$ (keV) &   ---     & ---&  --- &   $0.76\pm 0.01$  \\
$\mathrm{EM}_{\rm 0.8 keV}$\footnotemark[$*$] & ---     & --- &  --- & $(4.6\pm 0.2) \times 10^{-4}$   \\
Gaussian Energy (keV)   & --- &  0.561 (fix)& ---  & --- \\
Normalization\footnotemark[$\dagger$] of O\,\emissiontype{VII} He$\alpha$   & --- & $1.95\pm0.36$& --- & --- \\
Gaussian Energy (keV)   & --- &  $0.863\pm 0.007$& ---  & --- \\
Normalization\footnotemark[$\dagger$] of O\,\emissiontype{VIII} 6p to 1s? Fe?  & --- & $0.14\pm 0.01$& --- & --- \\
Gaussian Energy (keV)   & --- &  $0.929\pm 0.005$& ---  & --- \\
Normalization\footnotemark[$\dagger$] of Ne\,\emissiontype{IX} He$\alpha$?  Fe? & --- & $0.16\pm 0.01$& --- & --- \\
Gaussian Energy (keV)   & --- &  $1.040\pm 0.003$& ---  & --- \\
Normalization\footnotemark[$\dagger$] of Ne\,\emissiontype{X} Ly$\alpha$?  & --- & $0.11\pm 0.01$& --- & --- \\
Gaussian Energy (keV)   & --- &  $1.315\pm 0.006$& ---  & --- \\
Normalization\footnotemark[$\dagger$] of Mg\,\emissiontype{XI} He$\alpha$?  & --- & $0.063\pm 0.004$& --- & --- \\ \hline
C-Statistics/d.o.f & 6843/3524 & 5608/3515 & 5421/3520 & 5511/3522 \\   \hline
\end{tabular}}\label{tab:stacked}
\begin{tabnote}
\footnotemark[$*$] Emission measure ( integrated over the line of sight, $\int n_{\rm e}n_{\rm H} ds$) in units of $\rm{cm}^{-6}\rm{pc}$.\\
\footnotemark[$\dagger$] Normalization of Gaussian in 
units of photons $\rm{cm}^{-2}\rm{s}^{-1}\rm{sr}^{-1}$  (LU) \\
\footnotemark[$\ddagger$] Normalizations at 1 keV for the CXB component in 
units of photons $\rm{cm}^{-2}\rm{s}^{-1}\rm{keV}^{-1}\rm{sr}^{-1}$.\\ 
  \end{tabnote}	
\end{table*}

To reproduce the residual structures,
we added five Gaussians to Model-s.
Here, the line widths of these Gaussians were fixed at 0.
We fixed the line energy of a Gaussian at 0.561 keV,  which corresponds to  the forbidden line of  O\,\emissiontype{VII} He$\alpha$.
 Hereafter, we call this model Model-g.
The derived C-statistics,  5608,  is significantly better than that for Model-s.
This model can reproduce the observed line-like structures relatively well (table \ref{tab:stacked}, figure \ref{fig:stacked}). 
The best-fit $kT_{\rm halo}$, 0.22 keV, is close to those derived with Model-s.
The best-fit energies of the other additional Gaussians are
 0.86 keV, 0.93 keV, 1.04 keV, and 1.31 keV.
These energies are close to those of O\,\emissiontype{VIII}  6p to 1s line at 0.847 keV,  the forbidden line of  Ne\,\emissiontype{IX} He$\alpha$ at 0.905 keV,  Ne\,\emissiontype{X} Ly$\alpha$ at 1.022 keV, and  the forbidden line  of  Mg\,\emissiontype{XI} He$\alpha$ line at  1.331 keV,  respectively.

\citet{Yoshino09} and \citet{Nakashima18}
tried a variable abundance CIE model for the MWH component. They sometimes got a very high Ne abundance to explain the 0.9 keV peak in some spectra.
 We then replaced the {\it apec} model for the MWH component of the Model-s with {\it vapec} model (hereafter Model-v) and fitted the stacked spectra in the same way.
 The abundances of O, Ne, Mg, and Fe were allowed to vary, and those of the other elements were fixed at 1 solar.
 The result is shown in table \ref{tab:stacked} and figure \ref{fig:stacked}.
The derived C-statistics, 5421, is slightly better than that with Model-g. 
$kT_{\rm halo}$ increased to 0.33 keV, and the abundances of O, Ne, and Mg are around 0.8--0.9 solar, while the Fe abundance is only 0.2 solar.

\citet{Yoshino09} also tried a spectral fitting model with an additional CIE component with a temperature of 0.6--1 keV to reproduce the 0.9 keV peak.
Therefore,
 we added another {\it apec} component (hereafter 0.8 keV component) modified by the photoelectric absorption to  Model-s and fitted the spectrum.  
 Hereafter, we call this model  Model-08. 
  Here, the metal abundance was fixed at 1 solar.
  The derived C-statistics, 5511,  is also much smaller than that with Model-s.
  The best-fit temperature of the additional {\it apec} component is 0.76 keV, and $kT_{\rm halo}$ decreases to 0.18 keV.
As shown in figure \ref{fig:stacked}, Model-08 gives a better fit than Model-v below 1 keV, although it fails to produce the line-like structures at 1.0 keV and 1.3 keV, which correspond to the energy of Ne Ly$\alpha$ and  Mg He$\alpha$, respectively.

\subsection{Spectral fitting of the individual observations}

We then fitted the XIS spectra of the individual observations with Model-s and Model-08.
Since the Model-v fits for the individual observations give similar results to those obtained by \citet{Nakashima18}, we do not present the results in this paper.

\subsubsection{The standard model}

\begin{figure*}
\centerline{
   \includegraphics[width=8cm]{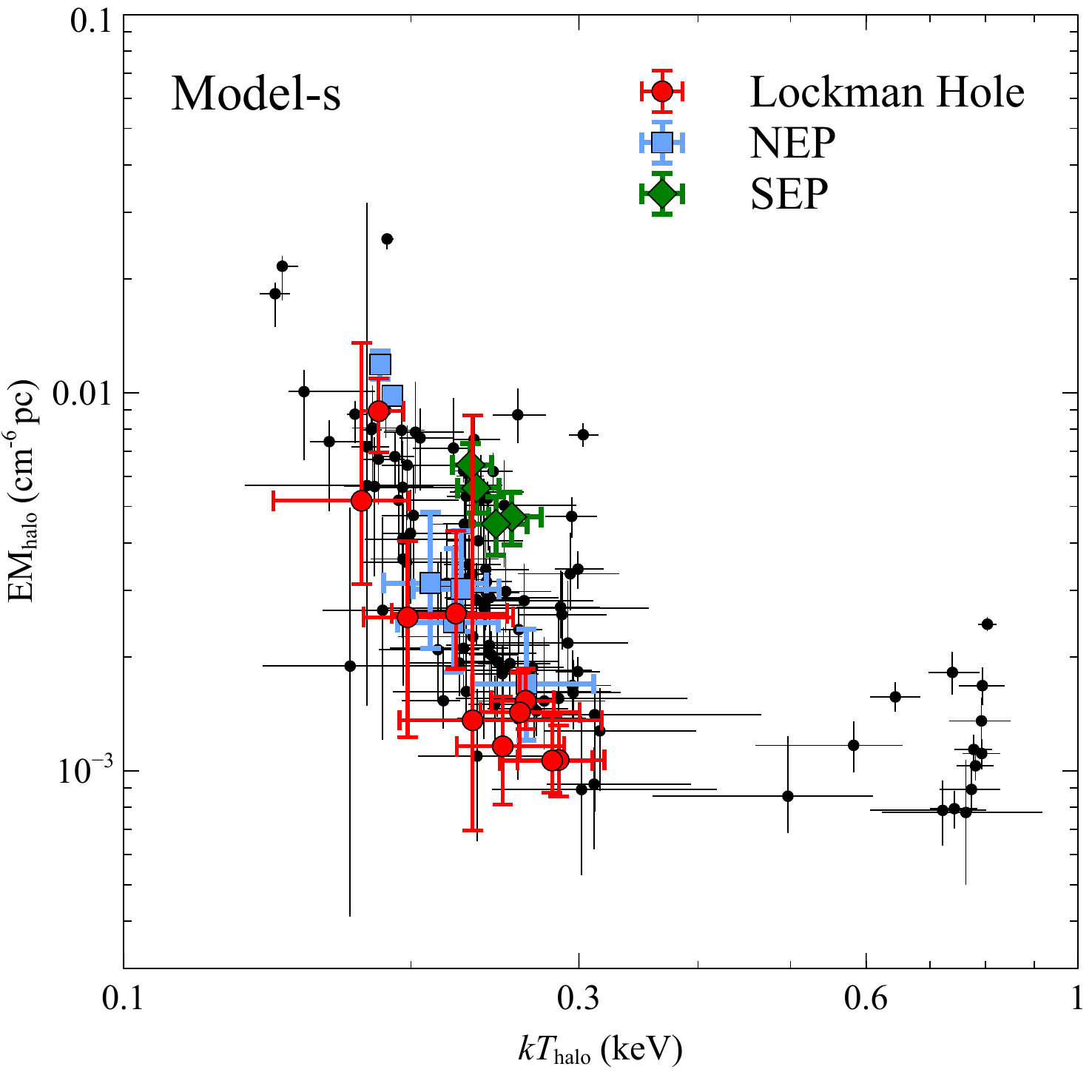} 
     \includegraphics[width=8.cm]{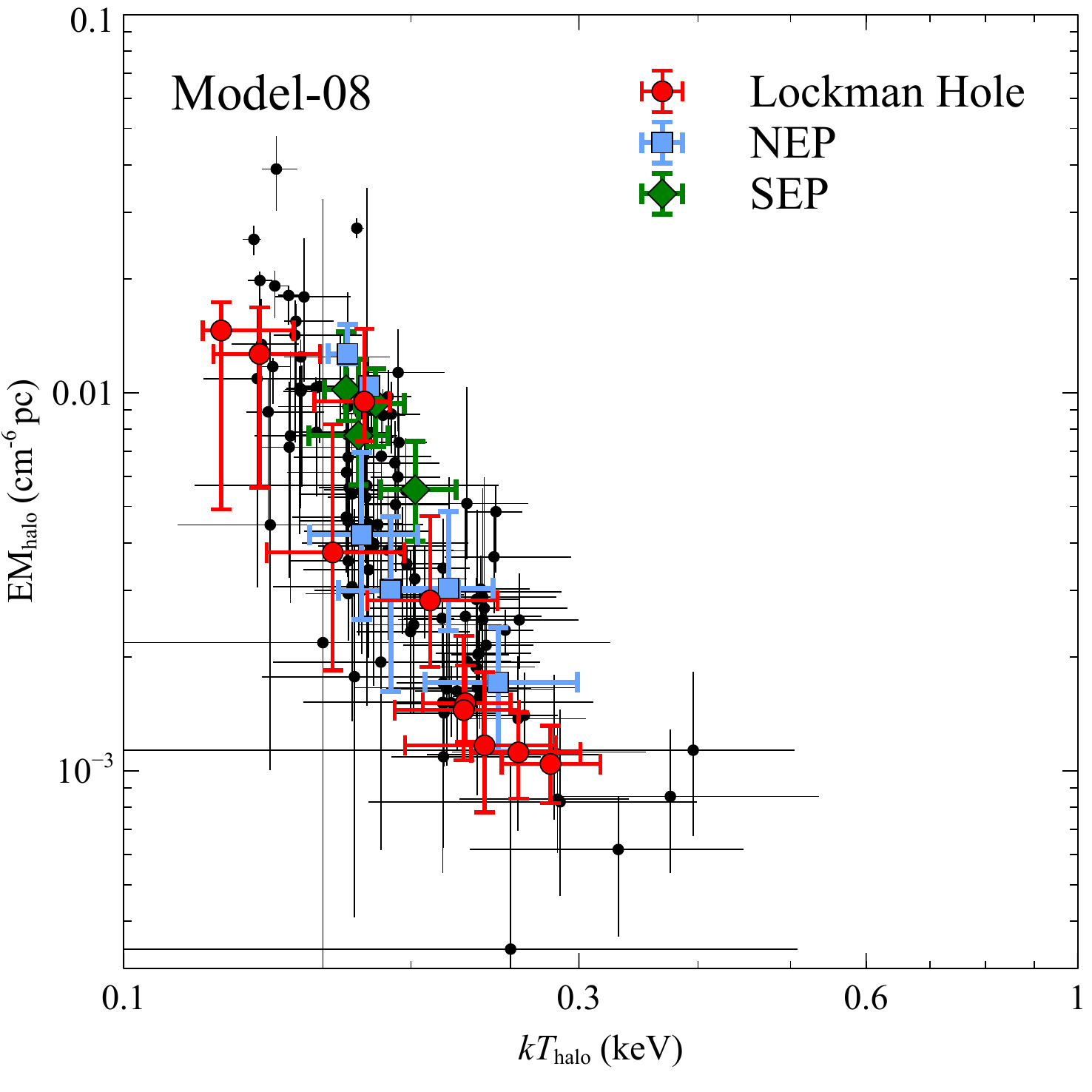} 
  }
\vspace*{1cm}
\caption{EM$_{\rm halo}$ against $kT_{\rm halo}$ with Model-s (left) and Model-08 (right).
The filled circles, filled squares, and filled diamonds correspond to the Lockman hole, NEP, and SEP, respectively.}\label{fig:EMkTModel-s}
\end{figure*}

 We fitted the XIS spectra of the individual observations with  Model-s.   Photoelectric absorption by cold gas in our Galaxy was modeled using {\it phabs} with a fixed hydrogen column density at the value by \citet{Kalberla05}.
The left panel of figure \ref{fig:EMkTModel-s} shows the scatter plot of
EM$_{\rm halo}$ vs. $kT_{\rm halo}$.
In most cases, $kT_{\rm halo}$ and EM$_{\rm halo}$ are in the range 0.15--0.3 keV and  (1--30)$\times 10^{-3}~\rm{cm^{-6}pc}$, respectively, and inversely correlated.
As reported by \citet{Yoshino09}, \citet{Sekiya14a}, and \citet{Nakashima18},
some observations show significantly higher $kT_{\rm halo}$ around  0.5--0.8 keV.
We find that observations with almost the same or nearby pointings show significantly different 
$kT_{\rm halo}$ and EM$_{\rm halo}$.
For example,  the nine Lockman Hole data obtained from 2006 to 2014 have almost the same sightlines.
However, their $kT_{\rm halo}$ and EM$_{\rm halo}$ are not consistent with each other and span similar ranges to those of the other observations.
EM$_{\rm halo}$ of the NEP observations in 2009, which point 1.2$^\circ$ offset from those in 2005 and 2006,  are significantly lower than those derived for NEP in 2005 and 2006.

\subsubsection{The spectral model with a hot component}

\begin{figure*}
\centerline{
    \includegraphics[width=8.cm]{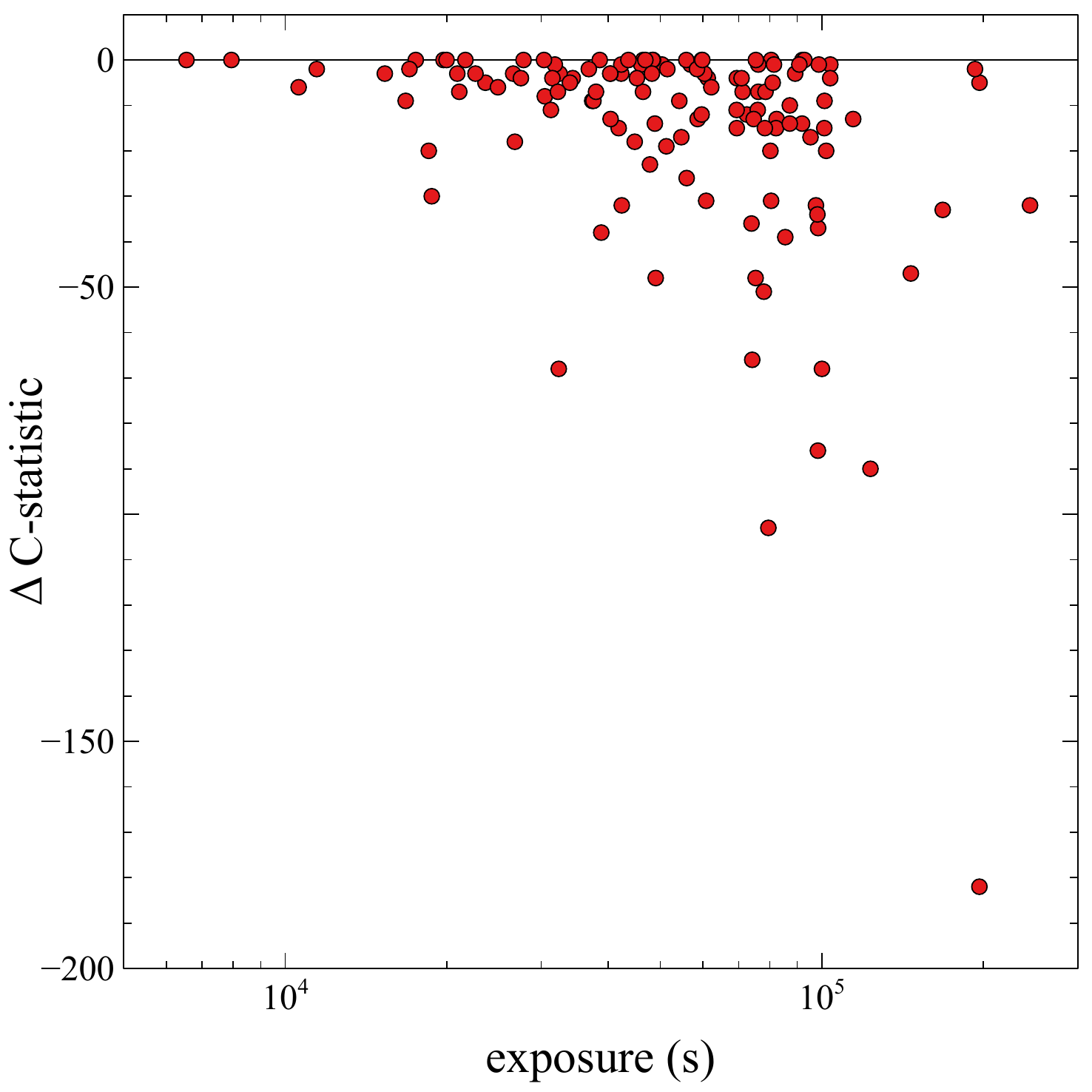} 
}
\bigskip
\caption{C-statistics for each observation derived from  Model-08 subtracted by that from Model-s,$\Delta$ C-statistic, plotted against exposure time.}
\label{fig:cstat}
\end{figure*}

We then fitted the spectra of each observation with Model-08 but fixed the temperature of the hottest {\it apec} component at 0.8 keV. 
As shown in figure \ref{fig:cstat},  Model-08 gives consistent or better fits than Model-s.
With Model-08, among 130 observations,  56 of them got $\Delta \rm{C}<-9$ compared to those with Model-s. Here, $\Delta$C is the difference in C statistic between the two models.
The results of the emission measure of the 0.8 keV component will be presented in Sugiyama et al. in preperation.
With Model-08, as shown in the right panel of figure \ref{fig:EMkTModel-s} and table \ref{tab:MWHmedian}, 
EM$_{\rm halo}$ spans an order of magnitude:
the median value is $4.0\times 10^{-3}~\rm{cm^{-6} pc}$ and the 16th-84th percentile range is  (1.6--10)$\times 10^{-3}~\rm{cm^{-6} pc}$.
The median and the 16th-84th percentile range of
$kT_{\rm halo}$ are 0.19 keV and 0.15--0.23 keV,  respectively.
$kT_{\rm halo}$ and EM$_{\rm halo}$ show a clear inverse correlation with a much smaller scatter than that for Model-s.
When we divide the sample into two subsamples with $kT_{\rm halo}=0.19$ keV, the median values of EM$_{\rm halo}$ differ by a factor of $\sim$ 4 (table \ref{tab:MWHmedian}).
Again,  those of the Lockman Hole observations with almost the same sightline span similar ranges to the other observations and are inconsistent with each other.

\begin{table*}
            \tbl{Medians and 16th-84th percentile ranges of MWH component with Model-08 }{
                      \begin{tabular}{cccccccc}\hline 
           selection     & N$^*$ &  \multicolumn{2}{c}{$kT_{\rm halo}$} &\multicolumn{2}{c}{EM$_{\rm halo}$}
           \\ & & median & 16th-84th percentile & median & 16th-84th percentile\\
               &    & (keV) & (keV) &  ($10^{-3}~\rm{cm^{-6}pc}$) & ($10^{-3}~\rm{cm^{-6}pc}$)\\
                 \hline 
                                 all & 130 & 0.19 & 0.15--0.23 & 4.0 & 1.6--10 \\
                   $kT_{\rm halo}^\dagger\ge$0.19 keV & 63 & --- & --- & 2.1 & 1.1-3.8 \\
	               $kT_{\rm halo}^\dagger<$0.19 keV& 67 &--- &--- & 7.9  & 3.9-13 \\
2005-2009 & 64 & 0.22 & 0.18-0.25 & 2.4 & 1.4-5.5\\
2010-2015 & 66 & 0.17 & 0.15-0.23 & 6.8 & 2.9-13\\	             
          SSN$^\ddagger$ $<$ 20 & 49 & 0.22 & 0.18-0.25 & 2.1 & 1.2-4.8 \\
                           20$\le$SSN$^\ddagger$ $\le$ 50 & 30 & 0.18 &  0.16-0.25 & 3.2 & 1.4-10\\
	             SSN$^\ddagger$ $\ge$ 50 & 51 & 0.17 & 0.15-0.22 & 7.7 & 3.1-15\\
	     \hline
            \end{tabular}}\label{tab:MWHmedian}
                \begin{tabnote}
                    \footnotesize
                    \footnotemark[$*$] Number of observations.\\
	               \footnotemark[$\dagger$] Temperature of the MWH component.\\
	               \footnotemark[$\ddagger$] 13 month-averaged sunspot number.\\
	                         \end{tabnote}
                \end{table*}

\subsection{Correlations with the solar activity}

\begin{figure*}
\centerline{
  \includegraphics[width=8cm]{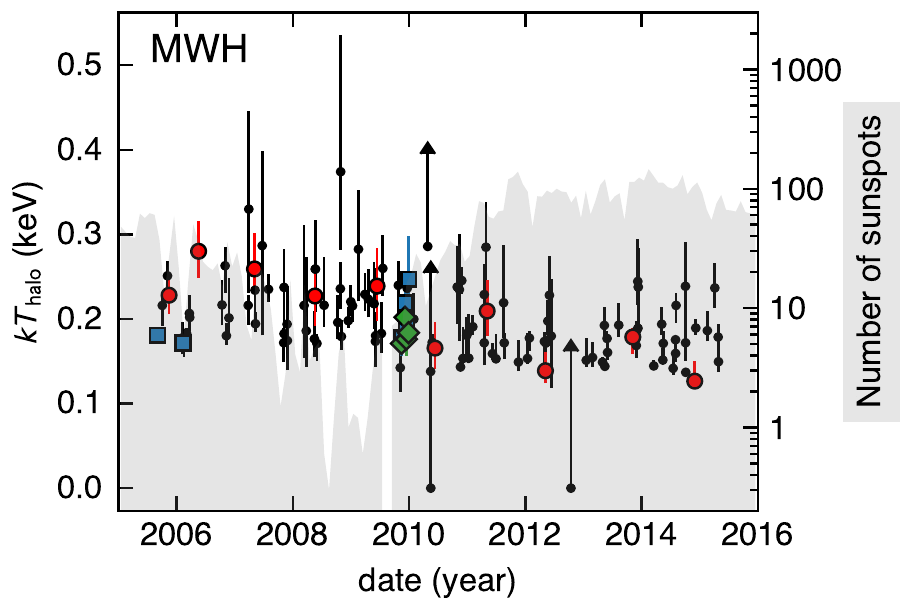}
  \includegraphics[width=8cm]{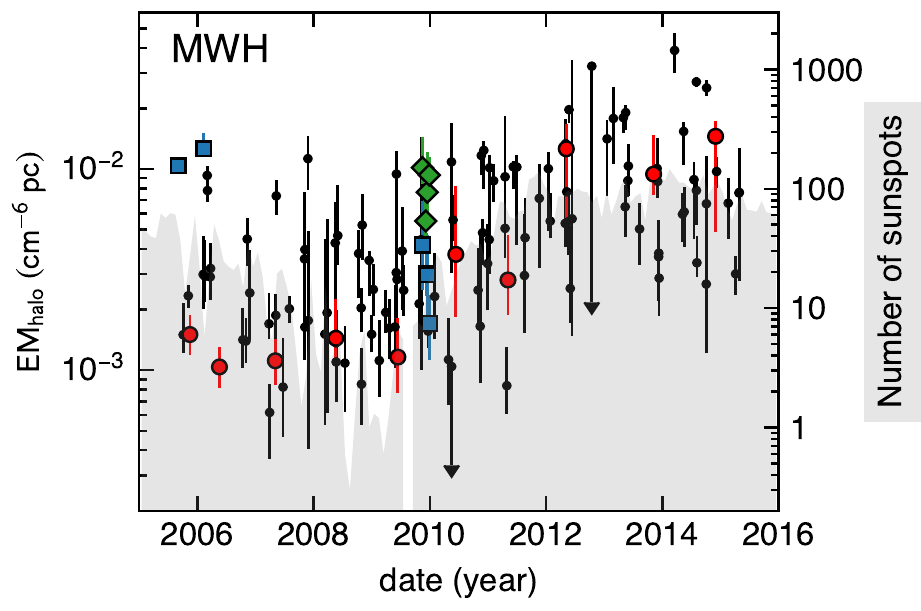}}
    \centerline{
  \includegraphics[width=8cm]{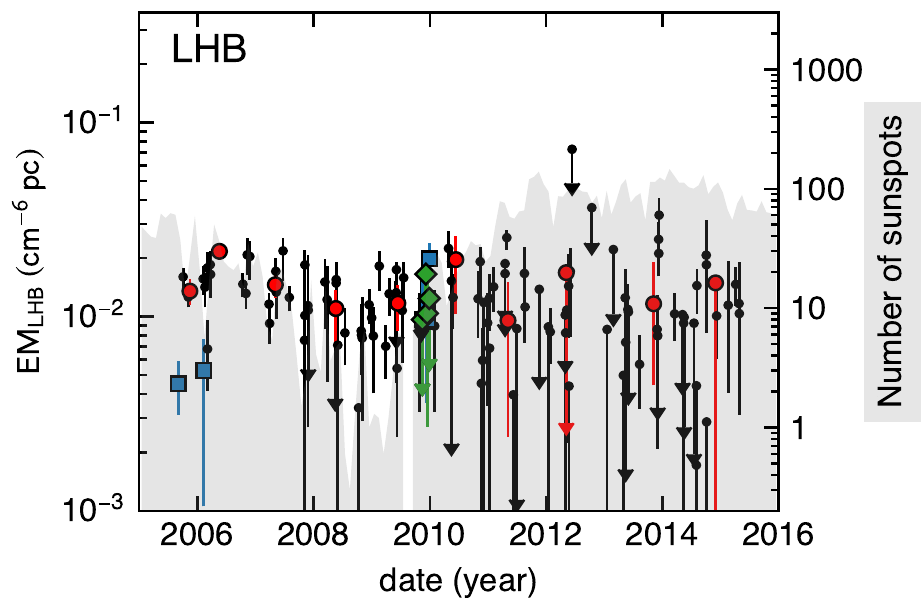}
  \includegraphics[width=8cm]{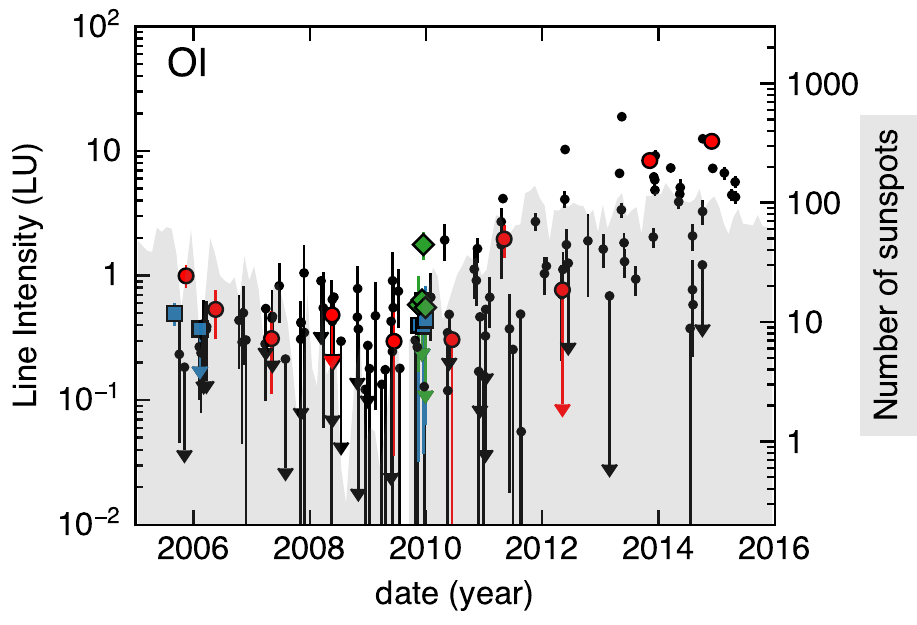}}
\caption{$kT_{\rm halo}$ and EM$_{\rm halo}$, the emission measures of the LHB and the normalization of O\,\emissiontype{I},  plotted against observation date.
Filled circles, filled squares, and filled diamonds correspond to the Lockman Hole, NEP, and SEP, respectively.  The gray shaded areas represent the 13-month averaged sunspot number.}\label{fig:date}
\centerline{
  \includegraphics[width=7cm]{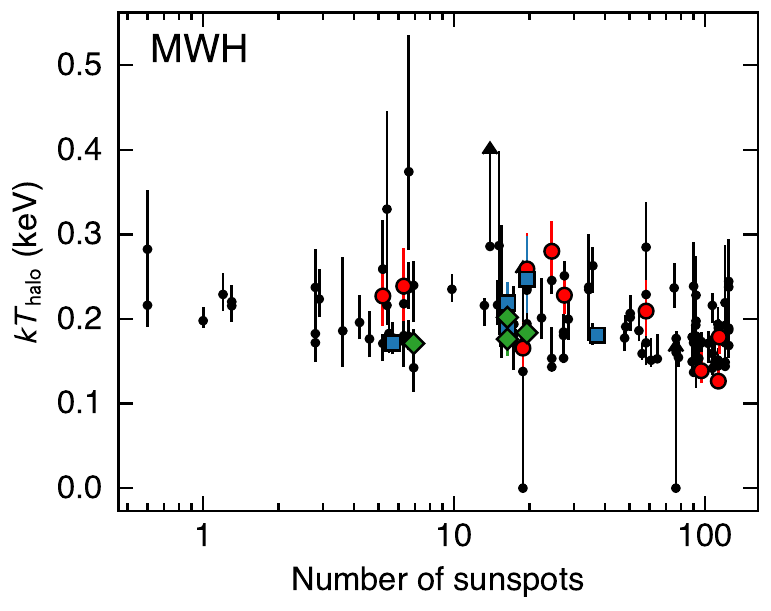}
  \includegraphics[width=7cm]{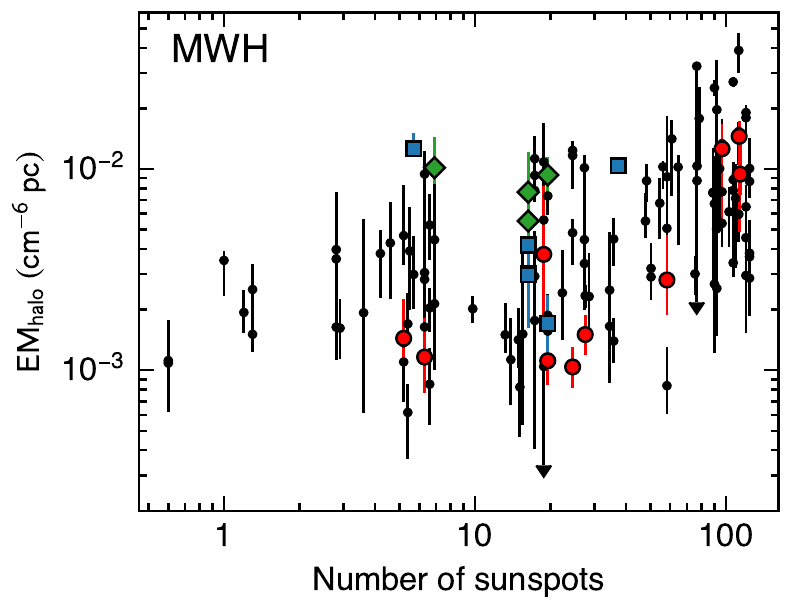}}
    \centerline{
  \includegraphics[width=7cm]{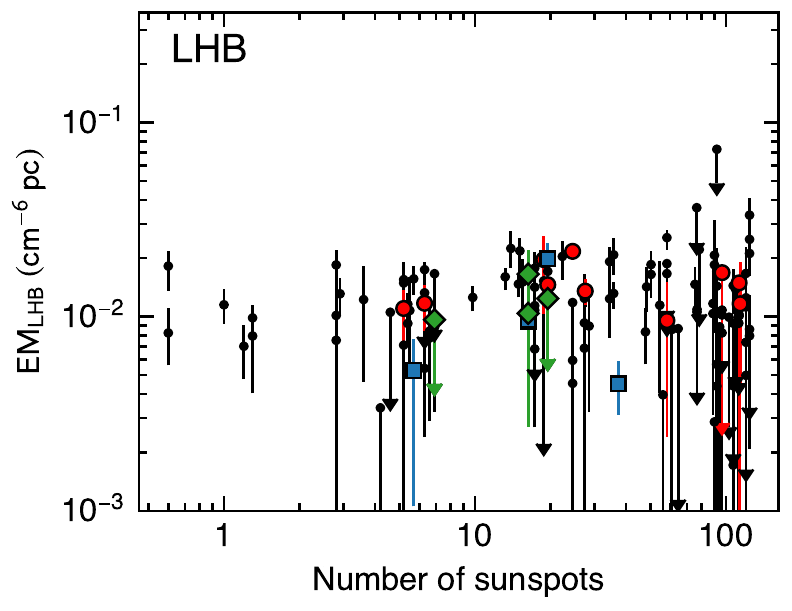}
  \includegraphics[width=7cm]{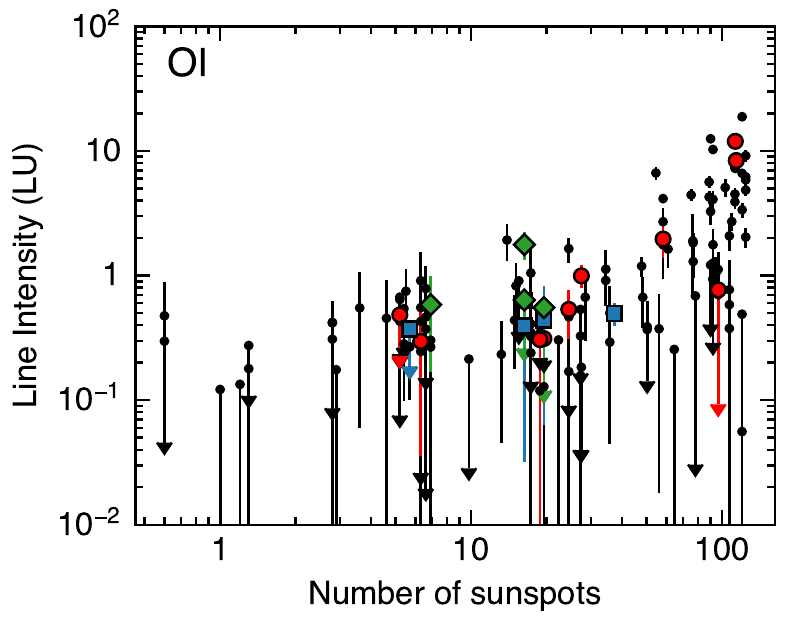}}
\caption{Same as Figure \ref{fig:date}, but plotted against 13-month averaged sunspot number.
}\label{fig:sunspot}
\end{figure*}

 Suzaku was operated from 2005 to 2015, covering nearly one solar cycle, including the solar minimum around 2009 and the solar maximum around 2014.
  \citet{Yoshitake13} found that the MWH emission measure depends on the solar activity using the Lockman hole data from 2006 to 2011 obtained with Suzaku.
The discrepancy among the observations with almost the same lines of sight indicates the existence of time-variable emission.  In figure \ref{fig:date}, we plot 
$kT_{\rm halo}$, EM$_{\rm halo}$, emission measures of LHB, and normalizations of the O\,\emissiontype{I} line with Model-08
against observation date with the 13-month smoothed sunspot number\footnote{$<$http://www.sidc.be/silso/dayssnplot$>$}.
$kT_{\rm halo}$ tend to be higher than 0.2 keV (the median value is 0.22 keV) before the end of 2009, while most of them are lower than 0.2 keV (the median value is 0.17 keV) after 2010.
As shown in figure \ref{fig:date}, the Lockman Hole data obtained in 2012-2015 also 
show much higher EM$_{\rm halo}$ than those obtained in 2006--2010. 
Most of EM$_{\rm halo}$  of the 130 observations  show a similar trend to those of the Lockman Hole. When we divide the data taken before and after the end of 2009 (hereafter 2005-2009 data and 2010-2015 data, respectively), the scatters are relatively small, as shown in table \ref{tab:MWHmedian}.

The time dependence of EM$_{\rm halo}$ resembles the 13-month smoothed sunspot number.
Therefore, in figure \ref{fig:sunspot}, we plot $kT_{\rm halo}$, EM$_{\rm halo}$, emission measures of LHB, and normalizations of the O\,\emissiontype{I} line with Model-08 against the 13-month smoothed sunspot number. The medians and the 16th-84th percentile ranges of different sunspot numbers are summarized in table \ref{tab:MWHmedian}.
For the data taken at the sunspot number is less than 20,  $kT_{\rm halo}$  and EM$_{\rm halo}$ are rather uniform, with their median values are 0.22 keV and $2.1\times 10^{-3}~\rm{cm^{-6} pc}$, respectively. 
In contrast, for data taken at the sunspot numbers are larger than several tens,  EM$_{\rm halo}$  increases dramatically.

As found by \citet{Sekiya14OI},  the brightness of the O\,\emissiontype{I} line  increases  after the solar minimum around 2009 (Figure \ref{fig:date}).
In contrast, 
the emission measure of the LHB  does not show a significant dependence on the observation date or 13-month smoothed sunspot number (figures \ref{fig:date} and \ref{fig:sunspot}).

\subsection{Time variable emission components with the Lockman Hole observations}

\begin{figure}
\centerline{  \includegraphics[width=8cm]{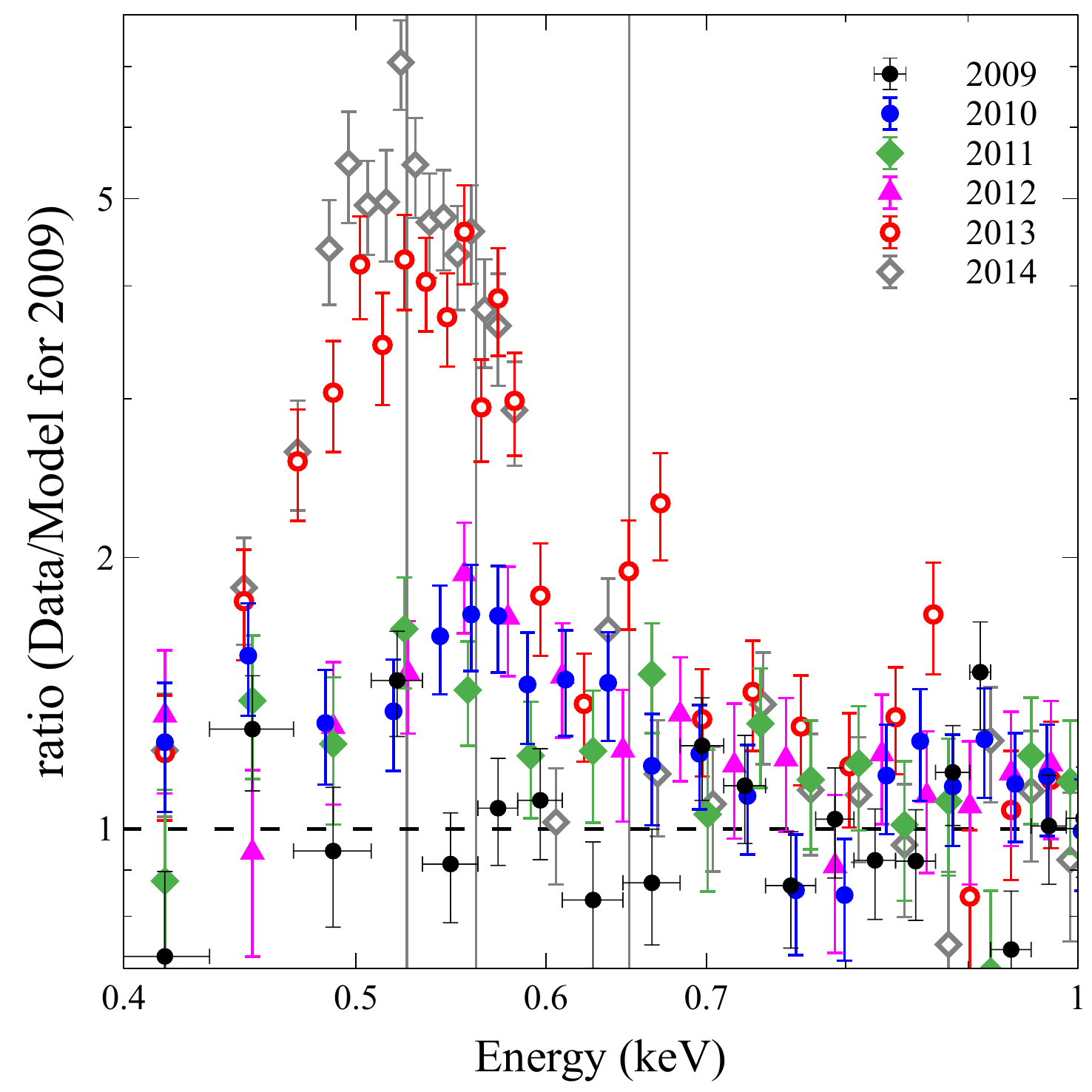}}
\caption{The ratios of the XIS1 spectra of the Lockman Hole obtained from 2009 to 2014 to the best-fit model for the 2009 observation with Model-08. The three vertical lines indicate the line energies of O\,\emissiontype{I}, O\,\emissiontype{VII} He $\alpha$ and O\,\emissiontype{VIII} Ly $\alpha$
}
\label{fig:lockman}
\end{figure}

  Figure \ref{fig:lockman} shows the ratios of the XIS1 spectra of the Lockman Hole obtained from 2009 to 2014 to the best-fit model of the 2009 observation with Model-08.
After the solar minimum around 2009, there is an enhancement at 0.56 keV which corresponds to O\,\emissiontype{VII} He$\alpha$. There is a weaker peak at O\,\emissiontype{VIII} Ly$\alpha$ (0.65 keV) in the 2013 spectrum. In addition, a strong peak at 0.525 keV of the O\,\emissiontype{I} line is seen in the spectra obtained in 2013.

We fitted the XIS1 spectra of Lockman Hole obtained from 2006 to 2014 simultaneously with Model-08, but we added two Gaussians at the fixed energies at O\,\emissiontype{VII} He$\alpha$ (0.56 keV) and O\,\emissiontype{VIII} Ly$\alpha$ (0.65 keV).
The normalizations of the additional Gaussians are fixed at zero for the 2009 spectrum while allowed to vary for the others.
The temperature of the MWH and 0.8 keV components were fixed at 0.22 keV and 0.80 keV, respectively and 
the normalizations of LHB, MWH, and the 0.8 keV component were assumed to have the same values.
This model yields a good fit with a C-statistic/d.o.f of 11937/11305.
Although there is a hint of excess at 
0.85 keV (O\,\emissiontype{VIII}  6p to 1s) in the spectra of 2013, when we added one more Gaussian to the model,  its significance is less than 2 $\sigma$.

\begin{figure}
\centerline{  \includegraphics[width=8cm,clip=]{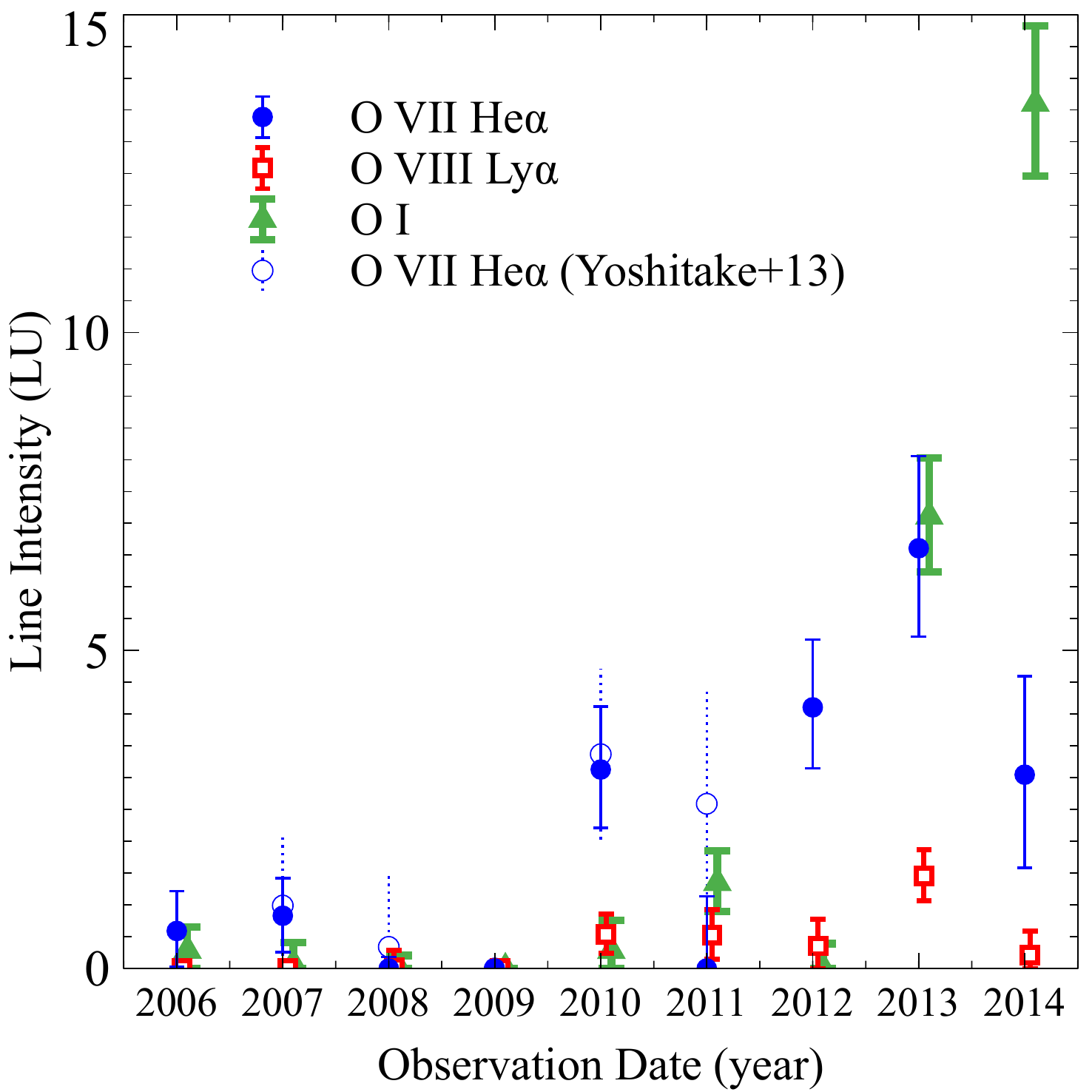}}
\caption{The normalizations of the Gaussians in units of LU (photons s$^{-1}$cm$^{-2}$str$^{-1}$) for the excess emissions of the Lockman Hole data compared to the 2009 data at 0.56 keV (filled circles, O\,\emissiontype{VII} He$\alpha$), 0.65 keV (open squares, O\,\emissiontype{VIII}  Ly$\alpha$), and 0.525 keV (filled triangles,O\,\emissiontype{I}).  The open circles with dotted error bars correspond to the excess O\,\emissiontype{VII} He$\alpha$ line strengths compared to the 2009 data obtained by \citet{Yoshitake13}. }
\label{fig:lockmanO}
\end{figure}

Figure \ref{fig:lockmanO} shows the excess line intensities of the O\,\emissiontype{VII} He$\alpha$, O\,\emissiontype{VIII} Ly$\alpha$, and O\,\emissiontype{I} compared to the 2009 spectrum plotted against the observation date.
The intensities of the excess O\,\emissiontype{VII} line increased from the solar minimum in 2009 to the solar maximum in 2014.
Figure \ref{fig:lockmanO} also compares the excess line strengths of the O\,\emissiontype{VII} line compared to the 2009 data obtained by \citet{Yoshitake13}.
Our excess O\,\emissiontype{VII} line strength for the  2010 observation agrees well with the value derived by \citet{Yoshitake13}, although our screening criteria are not severe.
In contrast, our excess O\,\emissiontype{VII} line strength for the 2011 data is lower than that obtained by \citet{Yoshitake13}, 
since their model did not include the O\,\emissiontype{I} line component.
The excess O\,\emissiontype{VIII} line strengths are much weaker than those of the O\,\emissiontype{VII} line.
We detected the O\,\emissiontype{VIII} line with $>$~3$\sigma$ significance only in 2013, where
 its intensity is about 20\% of that of the O\,\emissiontype{VII} line.
 As found by \citet{Sekiya14a}, the O\,\emissiontype{I} line is very bright around the solar maximum in 2014.

\section{Discussion}

\subsection{Emission related to the solar activity}

To study the soft X-ray background emission, we analyzed 130 Suzaku observations, covering nearly one solar cycle, including the solar minimum around 2009 and the solar maximum around 2014.
The overall trend of $kT_{\rm halo}$  and EM$_{\rm halo}$ with Model-08 for the 130 observations seems to depend on the solar activity (figures \ref{fig:date} and \ref{fig:sunspot}).
At a given epoch, the scatter in these values are relatively small, and most of the derived values are close to those obtained from the Lockman Hole observations.
Although we screened the possible geocoronal SWCX emissions using the light curve of each observation, screening the heliospheric one is challenging since the variation time scale is expected to be longer.
From the Lockman Hole data with almost the same sightlines, the excess the O\,\emissiontype{VII} 
He$\alpha$ emissions of 3--7 LU (photons s$^{-1}$cm$^{-2}$str$^{-1}$) are detected in 2010, 2012--2014 compared to the 2009 data. 
The intensities of the excess O\,\emissiontype{VIII} Ly$\alpha$ line are less than 20\% of those of O\,\emissiontype{VII} He$\alpha$.
The O\,\emissiontype{VII}/O\,\emissiontype{VIII} line ratios of the excess emission are significantly different from those of the geocoronal SWCX where the strengths of the OVIII line are comparable to those of the O\,\emissiontype{VII} line (e.g. \cite{Fujimoto07, Ishi19}).
\citet{Yoshitake13} estimated the heliospheric SWCX emission based on the model by \citet{Koutroumpa06}. The predicted intensities of O\,\emissiontype{VII} and O\,\emissiontype{VIII} lines around the solar maximum are 2.5 LU  and 0.8 LU, respectively, and around the solar minimum are 1.7 LU and 0.4 LU, respectively.

With the XMM data covering 10 years, \citet{Qu22} concluded that the average MWH fluxes are reduced from 10 LU to 5.4 LU for O\,\emissiontype{VII} and from 2.5 LU to 1.7 LU for O\,\emissiontype{VIII}. These reductions are consistent with the excess   O\,\emissiontype{VII} and  O\,\emissiontype{VIII} line strengths obtained from the Lockman Hole data around the solar maximum.
These results indicate that the heliospheric SWCX emission, mainly O\,\emissiontype{VII} He$\alpha$ line significantly contaminates the spectra obtained around the solar maximum.

\begin{figure}
    \centerline{
  \includegraphics[width=8cm]{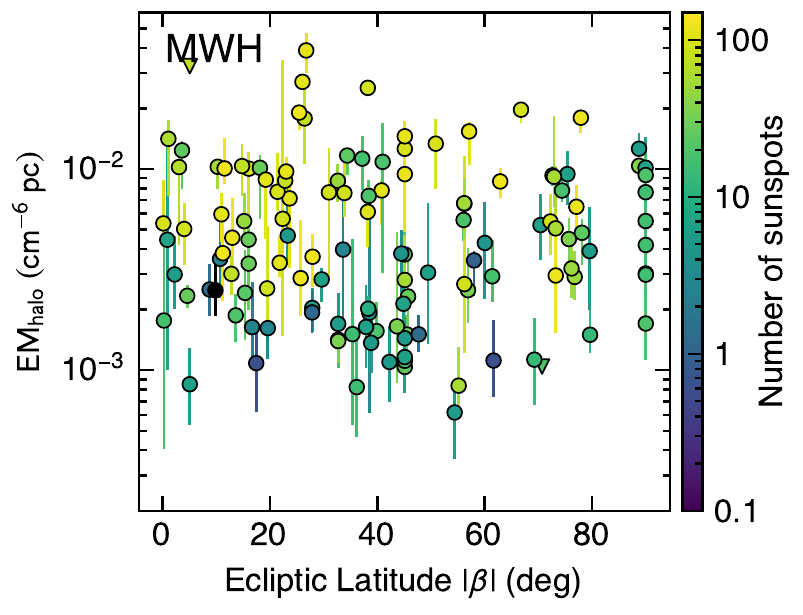}
 }
\caption{EM$_{\rm halo}$ plotted against the absolute value of the ecliptic latitude, $\beta$. 
The color scale indicates to the 13 months averaged sunspot number.
}\label{fig:mwhnormvsbeta}
\end{figure}

Figure \ref{fig:mwhnormvsbeta} shows EM$_{\rm halo}$ of the 130 observations plotted against the absolute value of the ecliptic latitude, with the color scale of the 13 months averaged sunspot number. 
The heliospheric SWCX is expected to be stronger near the 
ecliptic plane \citep{Robertson03,Koutroumpa06}. 
When the 13-month averaged  sunspot number is larger than several tens, EM$_{\rm halo}$ tends to be high.  However, no clear dependence on the ecliptic latitude is seen.
Probably, the time variation of the SWCX is much larger than the spatial variation along the ecliptic latitude.

 \subsection{Correlation between temperature and emission measure of the Milky way halo}

\begin{figure}
 \begin{center}
    \includegraphics[width=0.8\columnwidth]{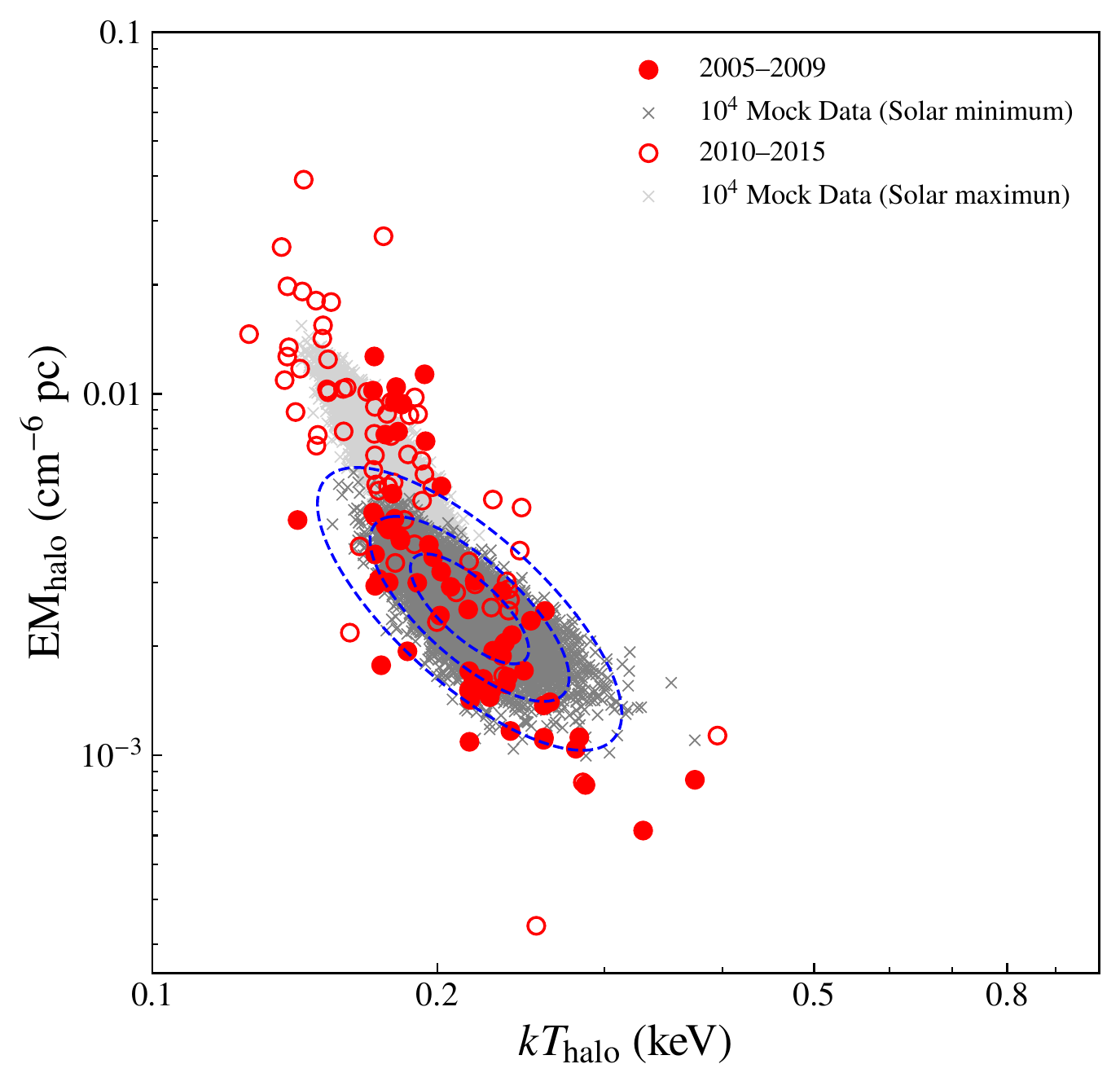} 
  \end{center}
 \bigskip
\caption{$kT_{\rm halo}$ plotted against  $EM_{\rm halo}$ (closed circles and open circles for the data of 2005-2009 and 2010-2015, respectively) with Model-08. The crosses (dark gray:2005-2009; light gray:2010-2015) shows the best-fit values The contours indicate the 68\%, 90\%, and 99\% ranges from the mock spectra using the median values for the 2005-2009 data.}\label{fig:ktnorm08corr}
\end{figure}

 The plot of EM$_{\rm halo}$--$kT_{\rm halo}$ shows a negative correlation as shown in  
 figure \ref{fig:ktnorm08corr}.
 Most of the 2010-2015 data give higher EM$_{\rm halo}$ and lower  $kT_{\rm halo}$ compared to the 2005-2009 data.
A CIE component with the temperature of $\sim 0.1$ keV has been empirically used to represent the emissions from  LHB and heliospheric SWCX (e.g.\cite{Yoshino09}, \cite{Nakashima18}).
However, 
 contaminations of the emissions related to the solar activity, mainly O\,\emissiontype{VII}  He$\alpha$, sometimes cause overestimation of  EM$_{\rm halo}$ and underestimation of $kT_{\rm halo}$, especially around the solar maximum.

  Since the emissivity of O\,\emissiontype{VIII} depends on the plasma temperature, 
to reproduce the observed O\,\emissiontype{VIII} line strengths, there may be some artificial negative correlation between EM$_{\rm halo}$--$kT_{\rm halo}$.
To check this effect, we created 10$^4$ mock XIS spectra with an exposure time of 50 ks  and the median values for the  2005-2009 and 2010-2015 data shown in table \ref{tab:MWHmedian} with Model-08.
Figure \ref{fig:ktnorm08corr} also shows the best-fit values from the mock spectra
with statistical contours.
Most of the best-fit values from the 2005-2009 data resemble the distribution from the mock spectra using their median values.
In contrast, the scatter for the 2010-2015 data points is larger than that from the corresponding simulations. 
  These results suggest that the negative correlation 
between EM$_{\rm halo}$ and $kT_{\rm halo}$ is likely artifacts and
most of EM$_{\rm halo}$ and $kT_{\rm halo}$ values of the 2005-2009 data are consistent with their median values.

\subsection{Comparison with previous results}

\begin{figure}
 \begin{center}
    \includegraphics[width=0.8\columnwidth]{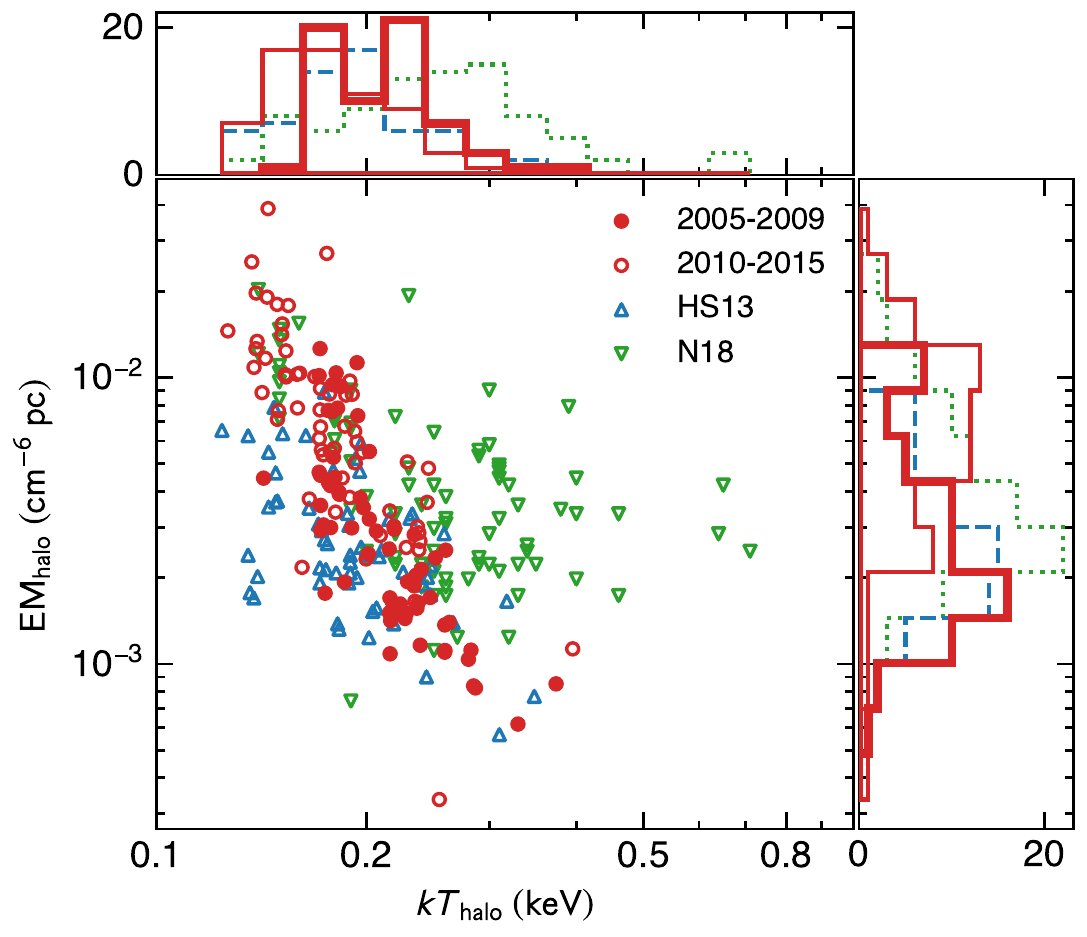} 
  \end{center}
 \bigskip
\caption{$kT_{\rm halo}$ plotted against  $EM_{\rm halo}$. The histograms show the distributions of $kT_{\rm halo}$ and $EM_{\rm halo}$. 
Closed circles and thick lines are for the 2005-2009 data and open circles and thin lines for the 2010-2015 data with Model-08, open upward triangles and dashed lines are for the data by \citet{Henley13}, and open downward triangles and dotted lines are for the data by \citet{Nakashima18}.  Here, the effect of the difference in the adopted solar abundance table is corrected.   }\label{fig:ktnorm08comp}
\end{figure}

Figure \ref{fig:ktnorm08comp} compares our $kT_{\rm halo}$ vs. EM$_{\rm halo}$ with those by \citet{Henley13} and \citet{Nakashima18}.
In order to minimize the contamination by SWCX, \citet{Henley13} analyzed only data after 2005 and $|\beta|>20^\circ$.
They also excluded data around the Orion-Eridanus superbubble \citep{Reynolds79, Burrows1993}.
As a result, their data do not include the highest emission measures which are possibly caused by the contamination of SWCX and by the emission from the bubble.
A significant fraction of our data, especially obtained before the end of 2009,  is consistent with that by \citet{Henley13}.
With the HaloSat survey data toward the southern halo ($b<-30^\circ$) obtained around the solar minimum, \citet{Kaaret20} studied the MWH halo. 
Their median temperature, 0.225 keV, agrees with our median value. Toward the Galactic anticenter, and high Galactic latitude, their emission measures, 4--8$\times 10^{-3}~\rm{cm^{-6} pc}$ (for 0.3 solar gas metallicity) agrees with our median values for the 2005-2009 data, considering the difference in the adopted abundance.

In our analysis, the screening criteria to minimize the SWCX is not as strict as in \citet{Nakashima18}. 
\citet{Nakashima18} used a similar variable abundance model with Model-v.
Over half of the data are consistent with our work.
The other data by \citet{Nakashima18} show higher $kT_{\rm halo}$  and  [O/Fe] $\sim$ 0.5 to reproduce
the residual structure at 0.7--1 keV, while we added the 0.8 keV component to the standard soft X-ray background model.
Then, the scatter in $kT_{\rm halo}$ vs. EM$_{\rm halo}$ is significantly reduced from those obtained by \citet{Nakashima18}.

It is reasonable to assume the MWH component extends over the Milky Way considering 
the small scatter in the $kT_{\rm halo}$ and EM$_{\rm halo}$  of 2005-2009 data,
and the residual structure at 0.7--1 keV seen in some regions may be caused by some additional component. The larger C-statistic from the Model-08 fit for the stacked spectra than the Model-v fit is caused by the excess Ne and Mg line emissions at 1 keV and 1.3 keV. There may be another higher temperature component ($>1$ keV) or the abundance pattern of the 0.8 keV component may deviate from the solar ratio.
  We will discuss these possibilities in a different paper (Sugiyama et al. in preparation).

\subsection{Spatial distribution of the Milky Way halo: the disk-like component}

\begin{figure*}
\centerline{
  \includegraphics[width=12cm]{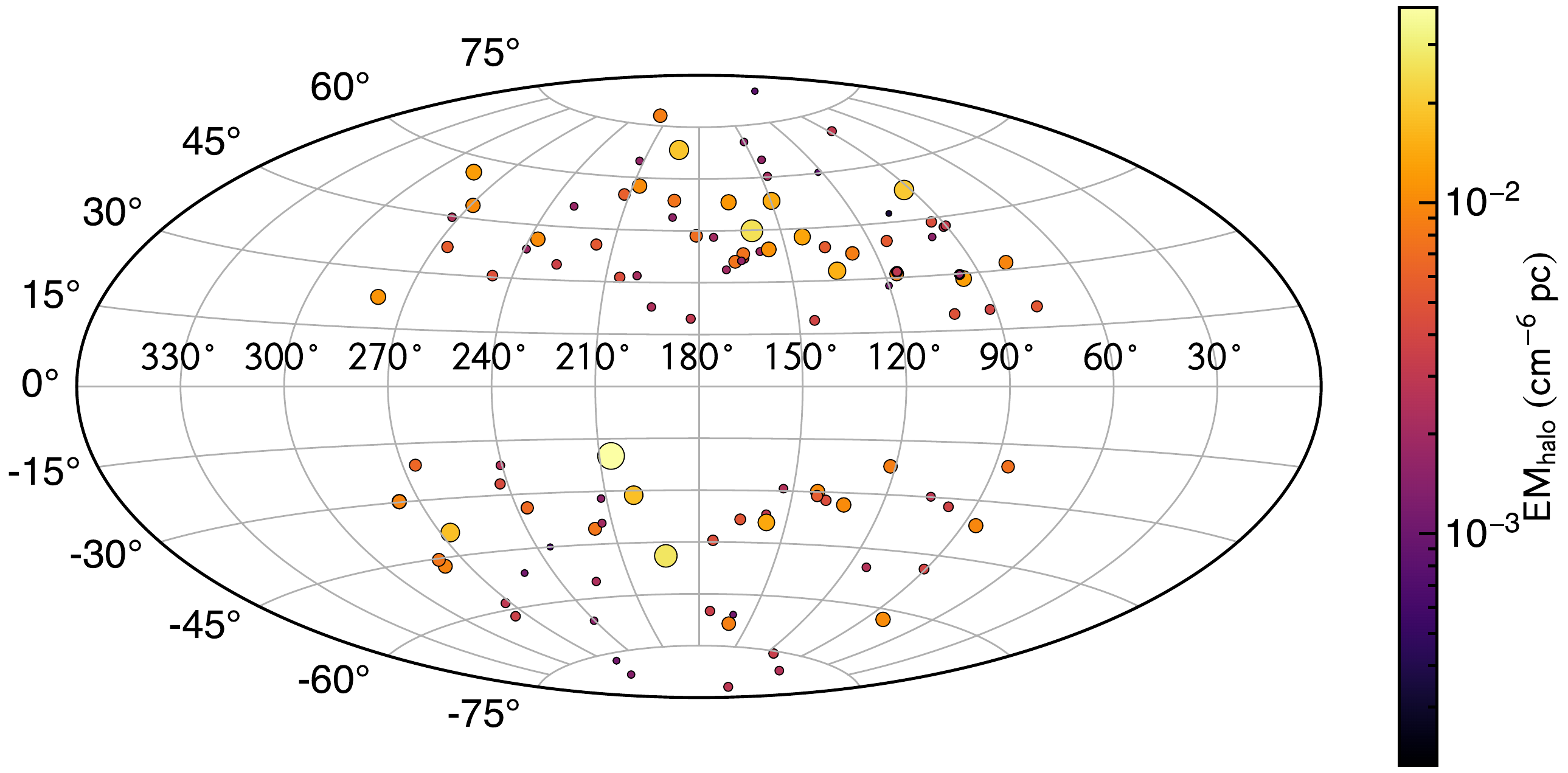} 
}
\caption{EM$_{\rm halo}$ across the sky with the galactic coordinates. The color and the size of circles indicate EM$_{\rm halo}$. }
\label{fig:galacticsky}
\vspace*{1cm}
\centerline{
  \includegraphics[width=8cm]{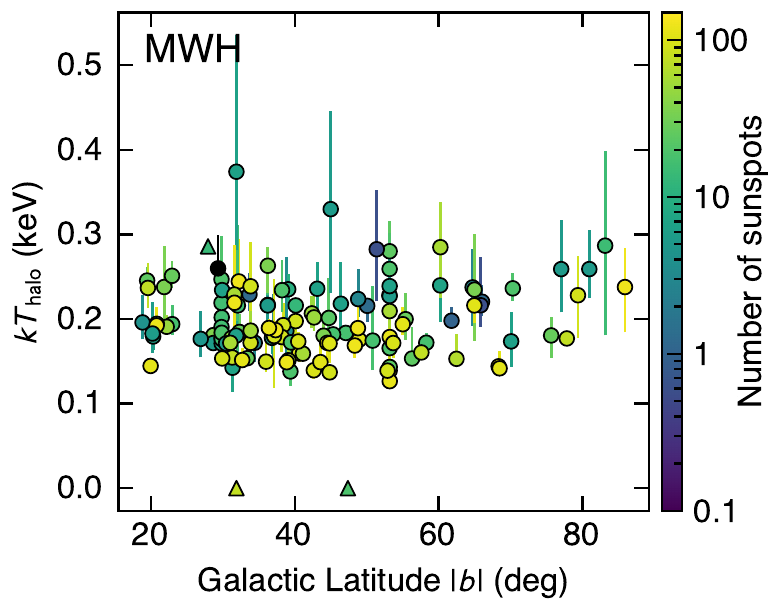}
  \includegraphics[width=8cm]{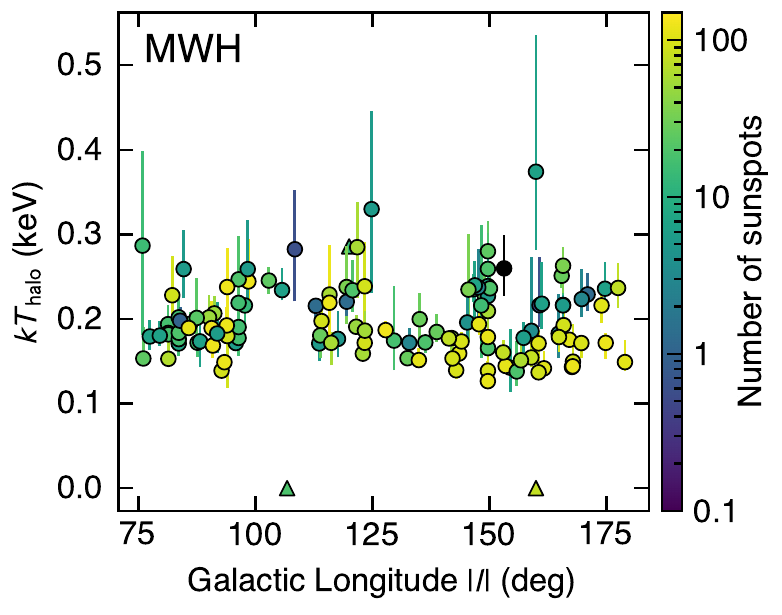}}
  \centerline{
  \includegraphics[width=8cm]{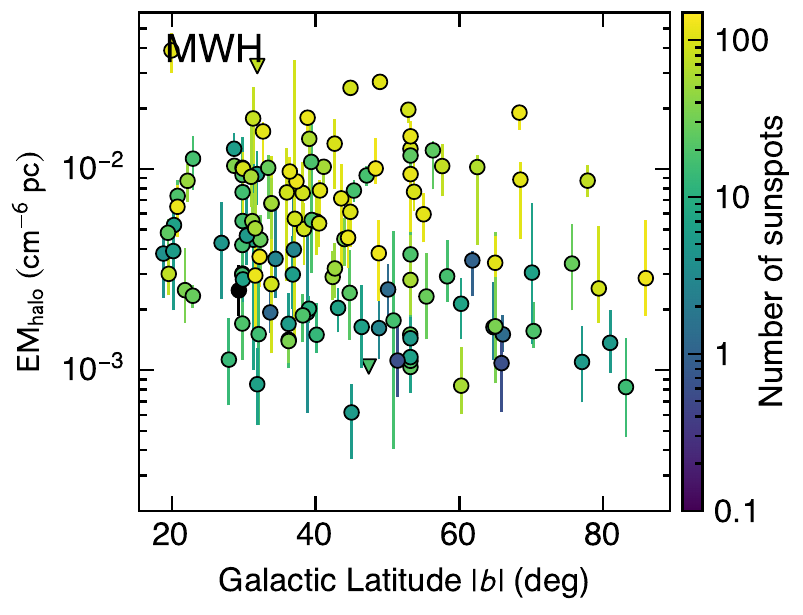}
 \includegraphics[width=8cm]{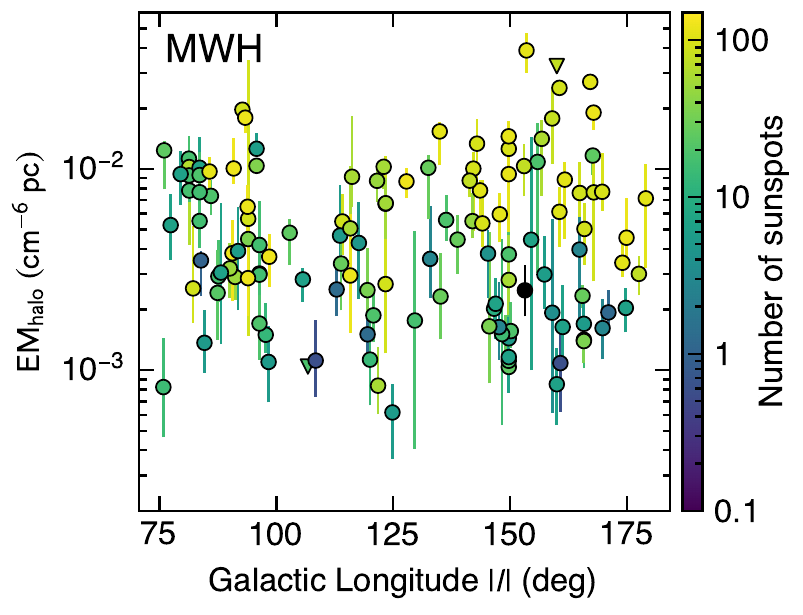}}
\caption{$kT_{\rm halo}$ and  EM$_{\rm halo}$ plotted against the absolute values of the  Galactic latitude, $|l|$ and the Galactic longitude, $|b|$.
The color scale corresponds to the 13-month-smoothed sunspot number.
}\label{fig:mwhnormvsb}
\end{figure*}

\begin{figure*}
   \includegraphics[width=16cm]{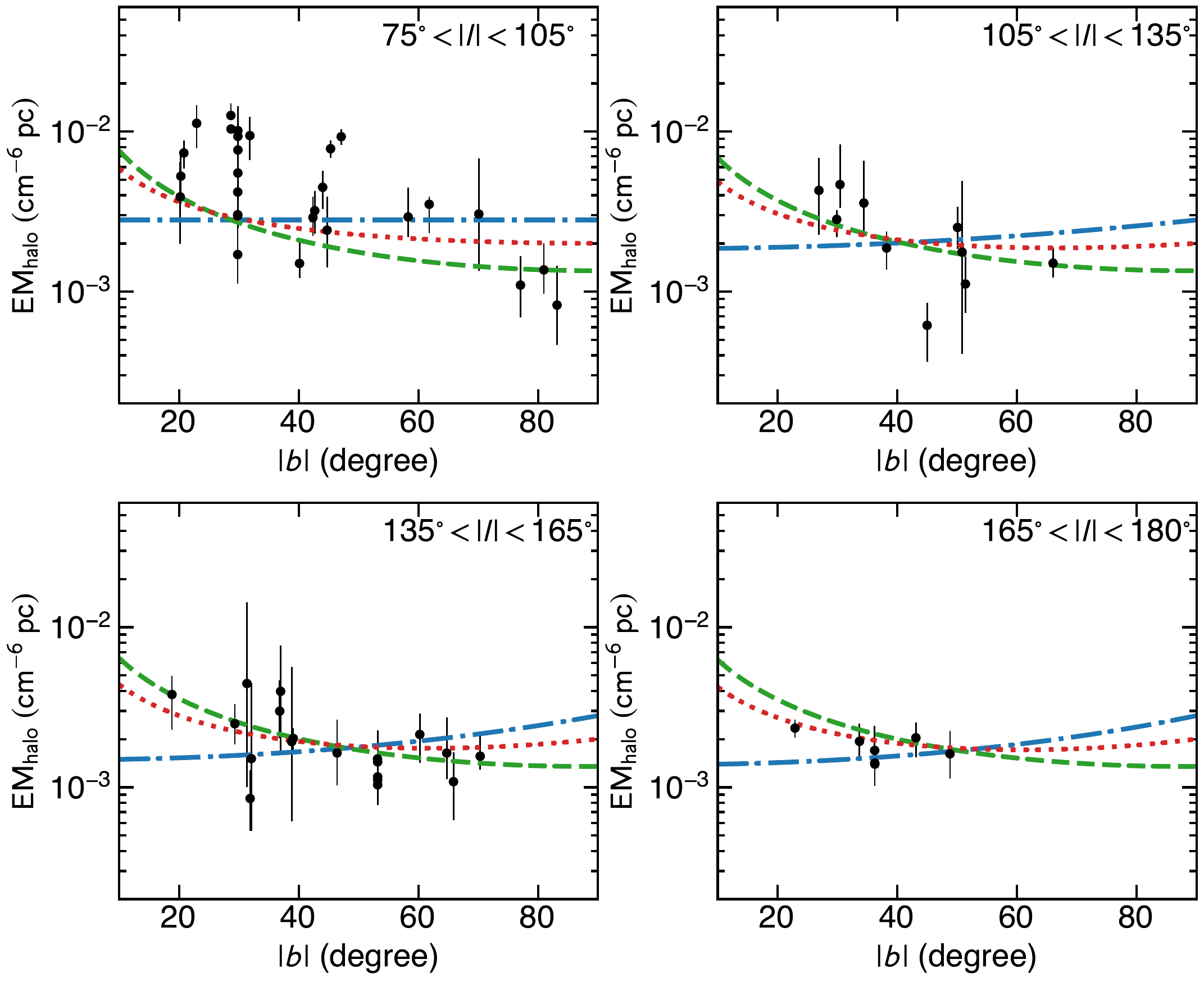}
 \caption{EM$_{\rm halo}$ for the 2005-2009 data,    plotted against the Galactic latitude, $|b|$ (filled circles with error bars). Each panel shows a different range of $|l|$. The dashed, dot-dashed, and dotted lines represent the best-fit disk-like, spherical, and composite models, respectively, at $|l|=90^\circ, 120^\circ, 150^\circ$, and $180^\circ$. }\label{fig:dist}
\end{figure*}

Figure \ref{fig:galacticsky} shows the sky map of the EM$_{\rm halo}$ with Model-08.
 Figure \ref{fig:mwhnormvsb} shows  $kT_{\rm halo}$ and  EM$_{\rm halo}$  with the same model fits, plotted against $|l|$ and $|b|$.
 Here, $|l|$ is defined as $l$ for $0^\circ\le l < 180^\circ$ and $360^\circ-l$ for $180^\circ\le l < 360^\circ$.
 The  brightest regions around $l\sim 200^\circ$ and $b\sim -20^\circ, -30^\circ,$ and $-50^\circ$  correspond to the Orion-Eridanus superbubble. 
Except for the regions with $|l|<105^\circ$ and $|b|<35^\circ$, 
$kT_{\rm halo}$ and  EM$_{\rm halo}$ are fairly uniform when we adopt the data with the sunspot number less than several tens.
Figure \ref{fig:dist} shows EM$_{\rm halo}$ for the 2005-2009 data, separated into four $|l|$ ranges.
For the data at $|l|>105^\circ$, the distribution of EM$_{\rm halo}$ is rather smooth, while there is a significant scatter for the data at  $|l|<105^\circ$.
For $|b|<35^\circ$ data, there is an increase of EM$_{\rm halo}$ toward the lower Galactic latitude 

The increase of the emission measure towards the low Galactic latitude indicates a presence of a disk-like morphology gas.
\citet{Yao09} proposed a disk-like morphology for the hot gas contributes the soft X-ray background as follows,
\begin{equation}
n_{\rm disk}=n_0\exp\left(-\frac{R}{R_0}\right)\exp\left(-\frac{z}{z_0}\right)
	\end{equation}
Here, $n_{\rm disk}$ is the number density, $n_0$ is the number density at
the Galactic center, $R$ is the distance from the Galactic center projected onto
the Galactic plane, $z$ is the vertical height from the Galactic disk, and $R_0$ and $z_0$ is 
the scale length and scale height, respectively.
The $n_0$ and $z_0$ obtained from previous absorption line studies
are about (1--5)$\times 10^{-3}~{\rm cm^{-3}}$, and 2--9 kpc \citep{Yao09, Hagihara2010,Sakai14}.
The HaloSat data toward the southern Galactic sky also prefers the disk-like morphology model, with $R_0=5.4\pm 1.5$ kpc, and $z_0=2.8\pm 1.0$ kpc \citep{Kaaret20}. 
We fixed $R_0$ and $z_0$ at 7.0 kpc and 2.7 kpc, respectively, which are the best-fit values obtained by \citet{Nakashima18}, and fitted the 2005-2009 data.
The results are summarized in table \ref{tab:dist}.
However, the fit is unacceptable with a $\chi^2/$d.o.f  is 389/63.
This large $\chi^2$ mainly comes from high EM$_{\rm halo}$ regions at $|l|<105^\circ$.
If we exclude the data with $|l|<105^\circ$, we obtain a good fit with $\chi^2$/d.o.f=36/35.
Since the O\,\emissiontype{VII} line contamination possibly from SWCX causes an underestimate of the temperature,
we fitted the data with $kT_{\rm halo}>0.2$ keV, including those obtained after 2010.
This fit also gives a relatively good fit with $\chi^2$/d.o.f=57/51. 
Assuming the solar metallicity, 
the central electron  density from the latter two fits is $(3.4\pm 0.1)\times 10^{-3}~\rm{cm}^{-3}$.  
However, it is difficult to measure the metal abundance of such low-temperature plasma. The derived electron density is roughly inversely proportional to the assumed metallicity. 
Integrating over this density out to 30 kpc, the gas mass becomes  $5\times 10^8(Z/0.3~ {\rm{solar}})^{-1}~ M_\odot$. 
Here, $Z$ is the metallicity of the hot gas.

This component may be related to the gas heated by supernovae explosions in the stellar disk.
Then, the hot gas may relate to star-forming regions, and the emission measures have some scatter.
The high EM$_{\rm halo}$ regions at $|l|<105^\circ$ may be caused by recent stellar feedback. However, 
the scale height of the SN-driven hot gas  ($T>10^{5.5}~\rm{K}$) from numerical simulation by \citet{Hill12}, about 0.3--0.5 kpc, is significantly lower
than the scale height of 2--9 kpc obtained from the absorption line studies.

\begin{table*}
\tbl{Fitting results of the density distribution models for MWH}{%
\begin{tabular}{clrrrccccccrr}\hline
 & & &\multicolumn{3}{c}{disk}&\multicolumn{2}{c}{spherical} &\\ 
selection       & model          & $N^*$   & ${{n_{\rm e}}_0}^{\dagger}$ & $z_0$ & $M(<30~\rm{kpc})^{\ddagger}$ & ${{n_{\rm e}}_{\rm c}}^\S$ &  $M(<250~\rm{kpc})^{\|}$ & $\chi^2$/dof \\
                &  &  &  ($10^{-3} ~{\rm cm}^{-3}$) &  (kpc)  & ($10^{8} ~M_\odot$)&$(10^{-3}~{\rm cm^{-3}})$ & ($10^{10}~ M_\odot$)\\
\hline
2005-2009 & disk &   64   & 3.9$\pm$0.2 & 2.7 & 5.6$\pm$0.3 & --- & --- & 389/63 \\ 
2005-2009, $|l|>105^\circ$ & disk & 36 & 3.4$\pm$0.1 & 2.7  & 4.9$\pm$0.2 & --- & --- & 36/35\\
$kT_{\rm halo}>0.2$ keV       & disk & 52 & 3.4$\pm$0.1 & 2.7&4.9$\pm$0.2& --- & --- & 57/51\\
2005-2009 & sphere &64  & --- & --- & --- & 4.1$\pm$0.2 & 4.2$\pm$0.2  & 439/63 \\
2005-2009,$|l|>105^\circ$& sphere &36  & --- & --- & --- & 3.6$\pm$0.2 & 3.7$\pm$0.2  & 59/35 \\
$kT_{\rm halo}>0.2$ keV       & sphere & 52 & --- & --- & --- & 3.5$\pm$0.1& 3.6$\pm$0.1 & 94/51\\
2005-2009 & comp &  64  & 5.9$\pm$2.4 &  0.3 & 1.0$\pm$0.4 &  2.6$\pm$0.9 & 2.7$\pm$0.9 &  408/62  \\
2005-2009,$|l|>105^\circ$ & comp &  36  & 7.8$\pm$1.7 &  0.3 &  1.3$\pm$0.3 &0.9$<$2.1&  0.9$<$2.2 &  32/34  \\
$kT_{\rm halo}>0.2$ keV       & comp  & 52 & 7.5$\pm$1.2 & 0.3 &1.2$\pm$0.2 & 1.1$\pm$0.8& 1.1$\pm$0.8&52/50\\
 \hline
\end{tabular}}\label{tab:dist}
 \begin{tabnote}
                    \footnotesize
                    \footnotemark[$*$] Number of observations.\\
	               \footnotemark[$\dagger$] Cental electron  density and scale height of the disk model,  assuming the metal abundance of 1 solar. \\
	               \footnotemark[$\ddagger$] Integrated gas mass of the disk component within 30 kpc, assuming the metal abundance of 0.3 solar\\
	               \footnotemark[$\S$] Cental electron  density of the spherical model,  assuming the metal abundance of 1 solar. \\              
	               \footnotemark[$\|$] Integrated gas mass of the spherical component within 250 kpc, assuming the metal abundance of 0.3 solar.\\
	                 \end{tabnote}
\end{table*}

\subsection{Spatial distribution of the Milky Way halo: the spherical component}

Galaxies are thought to be surrounded by a gaseous corona, the CGM.
The CGM is expected to be consists of mostly hot diffuse gas and contains
a significant fraction of baryonic mass.
This gas is expected to be heated to the virial temperature and fills the MWH in nearly hydrostatic equilibrium.
Numerical simulations indicate that the mass of the CGM 
is around  several percent of the total mass and for the Milky Way, where the expected CGM mass is about (3--10) $\times 10^{10} M_\odot$ (e.g. \cite{Hani19, Cautun19,F2022}).

The distribution of a hot gas filling a self-gravitating isothermal sphere in hydrostatic equilibrium is often modeled as the $\beta$-model,
\begin{equation}\label{eqss}
n_{\rm sphere}=n_{\rm c}\left(1+\left(\frac{r}{r_{\rm{c}}}\right)^2\right)^{-3\beta/2}$$.	
\end{equation}
Here, $n_{\rm sphere}$ is the number density of the gas, $n_{\rm c}$ is the core density, $r$ is the distance from the Galactic center, and $r_{\rm c}$ is the core radius.
$\beta=T_{\rm rot}/T$, where $T$ is the plasma temperature and $T_{\rm rot}$ is from the rotational velocity. 
Adopting the case with the optically thin O Ly$\alpha$, the absorption line measurements of our Galaxy indicate $\beta\sim 0.5$ \citep{Miller15,Bregman18}.
Since the Galaxy's rotation speed corresponds to 0.12 keV,  the virial temperature becomes 0.24 keV with $\beta=0.5$.
This temperature is close to the median value of $kT_{\rm halo}$, 0.22 keV,  with Model-08 for the 2005-2009 data.
Since our data is not enough to constrain $\beta$ and $r_{\rm c}$ from the spatial distribution of EM$_{\rm halo}$, we fixed them at
0.51 and 2.4 kpc, respectively, following \citet{Li2017}.
Then, integrating emission measure up to the virial radius (250 kpc, \cite{Cautun19}), we fitted the observed EM$_{\rm halo}$ of the 2005-2009 data with equation \ref{eqss}.
The results are shown in table \ref{tab:dist}.
Assuming the solar metallicity, 
the best-fit central  electron density is $(4.1\pm0.2)\times 10^{-3}\rm{cm^{-3}}$. The integrated gas mass out to the virial radius becomes 4.2$\times 10^{10}~ (Z/0.3 ~{\rm{solar}})^{-1}~M_\odot$. Therefore, if the gas metallicity is low, its mass  becomes comparable to the expected CGM mass,
5--6$\times 10^{10} ~M_\odot$ \citep{Cautun19}.
However,  this model yields $\chi^2/d.o.f=439/63$ and cannot reproduce the data at $|b|<40^\circ$ and some data at $|l|<105^\circ$.
When we exclude the data at $|l|<105^\circ$ from the 2005-2009 data or using the data with $kT_{\rm halo}>0.2 $ keV, the spherical model fit gives $\chi^2/d.o.f=59/35$ and $\chi^2/d.o.f=94/51$, respectively. These $\chi^2$ values are much larger than those obtained from the disk-like model fits. 

\subsection{Spatial distribution of the Milky Way halo: the composite model}

\begin{figure*}
    \includegraphics[width=14cm]{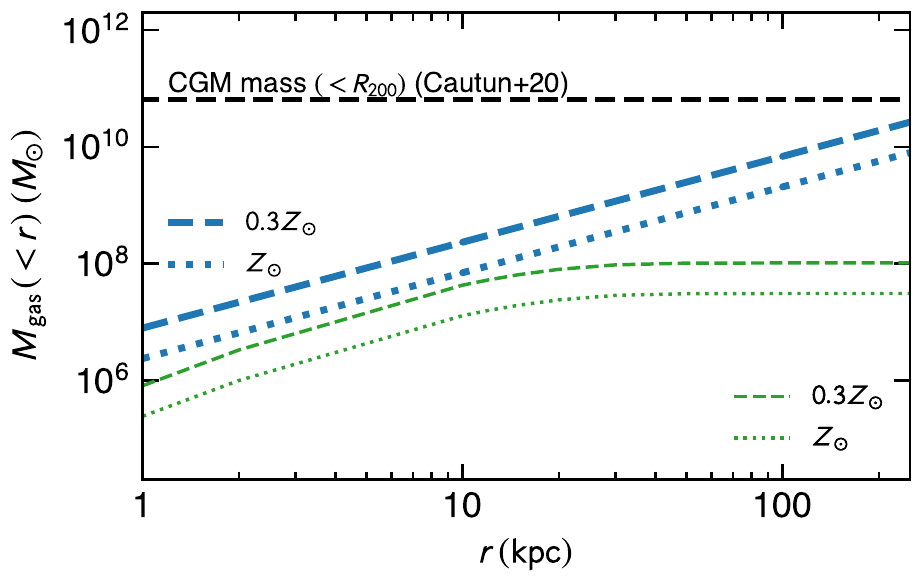}
\caption{The integrated gas mass (thick: spherical, thin: disk-like) with the best-fit composite model using the 2005-2009 data. The dashed and dotted lines correspond to the 0.3 solar and 1 solar abundance, respectively. The horizontal dashed line shows the expected CGM mass within $R_{200}$.}\label{fig:gasmass}
\end{figure*}

We then tried a composite model which consists of a disk-like and spherical gas distribution.
We assume the density described as
\begin{equation}
	n_{\rm comp}=n_{\rm disk}+n_{\rm sphere}
\end{equation}
In this model, we assume $\beta=0.51$, $r_{\rm c}=2.4$ kpc, and $R_0=7.0$ kpc.
We adopt the scale height $z_0$ at 0.3 kpc.
This scale height is consistent with the value expected by numerical simulations by \citet{Hill12}.
As shown in figure \ref{fig:dist} and table \ref{tab:dist},
this model also gives a good fit to the 2005-2009 data, except for data with $|l|<105^\circ$. 
Assuming the solar metallicity, 
the best-fit central electron  density of the disk and sphere components are $(5.9\pm 2.4)\times 10^{-3} ~{\rm cm^{-3}}$ and $(2.6\pm 0.9)\times 10^{-3} ~{\rm cm^{-3}}$, respectively.
Figure \ref{fig:gasmass} shows the integrated gas mass of the disk and spherical components, assuming 0.3 solar and 1 solar metallicity.
The integrated gas mass out to 30 kpc for the disk component and 250 kpc for the spherical component 
 are $(1.0\pm 0.4) \times 10^{8} (Z/0.3~ {\rm{solar}})^{-1}~M_\odot$ and $(2.7\pm 0.9) \times 10^{10} (Z/0.3~ {\rm{solar}})^{-1}~M_\odot$  ,  respectively.
The latter is comparable to the lower limit of the expected CGM mass.
When using the data at $|l|>105^\circ$ of the 2005-2009 data, or using the data with $kT_{\rm halo}>0.2$ keV, 
we get a good fit to the data with the reduced $\chi^2\sim 1$ (table \ref{tab:dist}).
In these two fits, the gas mass of the spherical component decrease from the previous fit, and its 
 upper limit out to 250 kpc is  $\sim 2\times 10^{10} (Z/0.3~ {\rm{solar}})^{-1}~M_\odot$.
Even if the spherical component extends out to this radius, most of the emission comes from a distance within a few tens of kpc.

\normalsize

\begin{figure*}
   \includegraphics[width=14cm]{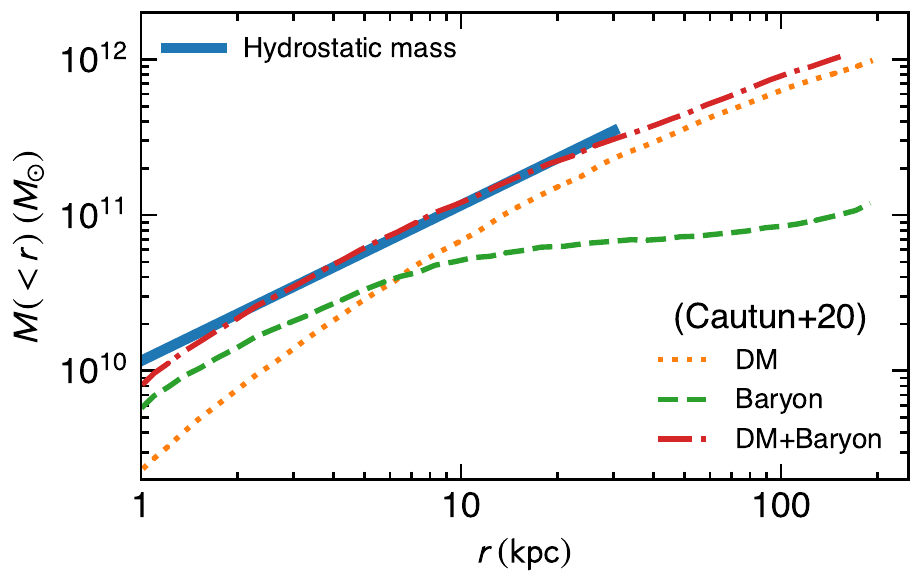}
\caption{The integrated hydrostatic mass from the spherical component (solid line). The integrated dark matter(dotted), baryon (dashed), and the sum of dark matter and baryon (dot-dashed) obtained by the Gaia observations \citep{Cautun19} are also shown.  }\label{fig:HEmass}
\end{figure*}

If gas is in hydrostatic equilibrium, the gravitational mass is given by
\begin{equation}
	M(r)=-\frac{kTr}{\mu m_{\rm p} G}\left(\frac{d\ln{n}}{d\ln{r}}+\frac{d\ln{T}}{d\ln{r}}\right)
\end{equation}
Here, $M(r)$ is the gravitational mass within $r$, $n$ is the hot gas density, $k$ is the Boltzman constant, $T$ is the temperature of the gas, $\mu$ is the mean molecular weight of 0.61, $m_{\rm p}$ is the proton mass, and  $G$ is the gravitational constant.
We adopted the temperature of 0.22 keV and $\beta$=0.51 and calculated the gravitational mass, assuming that the sphere component is in the hydrostatic equilibrium.
Figure \ref{fig:HEmass} shows the gravitational mass (hydrostatic mass) calculated out to 30 kpc, beyond which the contribution to the emission measure is minor.
The derived hydrostatic mass agrees very well with the total mass of dark matter and baryon obtained by Gaia observations \citep{Cautun19}.

The accreting gas from outside galaxies is likely in hydrostatic equilibrium, since
the expected accretion velocity is low (e.g. \cite{Bregman18}).
Our $kT_{\rm halo}$ is fairly uniform at the virial temperature.
The smooth distribution of EM$_{\rm halo}$ toward the Galactic anticenter and consistency of the hydrostatic mass and gravitational mass indicate that the hot gas with the virial temperature smoothly fills the halo of the Milky Way, possibly in hydrostatic equilibrium as theoretically expected.
However, because of the presence of the disk-like morphology component, it is challenging to constrain the contribution of the spherical component.

 \section{Summary}
 We analyzed data of 130 Suzaku
observations at $75^\circ<l < 285^\circ$ and $|b|>15^\circ$ obtained from 2005 to 2015, covering nearly one solar cycle.
We exclude time ranges with high count rates to minimize the effect of the geocoronal SWCX.
The standard soft X-ray background model consisting of the local hot bubble and the Milky Way Halo (MWH) fails to reproduce a significant fraction of the spectra.
We include an additional hot CIE component with a temperature of  $\sim 0.8$ keV to reproduce spectra of a significant fraction of the lines of sight.
Then, the scatter in the relation between the EM$_{\rm halo}$ vs. $kT_{\rm halo}$ is reduced.  However, the spectra of almost the same lines of sight, for example, the Lockman Hole data, are inconsistent. 
We simultaneously fitted the nine Lockman Hole spectra obtained from 2006 to 2014 and found an excess O\,\emissiontype{VII} He$\alpha$ emission after the solar minimum around 2009.
Excess O\,\emissiontype{VIII} Ly$\alpha$ is also detected at the solar maximum.
These results indicate
the heliospheric SWCX emission likely contaminates the spectra and causes underestimation of $kT_{\rm halo}$ and overestimation of EM$_{\rm halo}$.
Adopting the data taken before the end of 2009,
at $|b|>35^\circ$ and $|l|>105^\circ$, 
the temperature ($\sim$ 0.22 keV) and emission measure ($2\times 10^{-3}~\rm{cm^{-6}~pc}$) of the MWH are fairly uniform, while at $|b|<35^\circ$ the emission measure increases toward the lower Galactic latitude.
Toward the Galactic anticenter region ($|l|>105^\circ$), the 2005-2009 data are well represented with a disk-like morphology model and a composite model consisting of a disk and spherical morphology components.
The temperature, 0.22 keV, agrees with the virial temperature of the Milky Way.
The hydrostatic mass of the spherical component at a few tens of kpc from the Galactic center agrees with the gravitational mass of the Milky Way.
These results indicate that the plasma with the virial temperature fills the Milky Way halo with nearly hydrostatic equilibrium.
Integrating the density profile of the spherical component out to 250 kpc, or the virial radius, the upper limit of the gas mass of  a few$\times 10^{10} M_\odot$, assuming 0.3 solar metallicity of the gas.
This value is comparable to the lower limit than the expected mass of CGM.

\appendix
\chapter{Observation log}
\label{appendix:observation_info}

Suzaku observations used are listed in table \ref{tab:observation_info}.

\begin{longtable}{ccrrrcr}
    \caption{Suzaku Observation log}
    \label{tab:observation_info} \\
    \hline
    Sequence$^*$ & Target Name  & \(l\) {$^\dagger$} & \(b\) {\(^\ddagger\)} & \(N_{\rm H}\) {\(^\S\)} & date {\(^\|\)} & Exposure Time (ks) \\ \hline
    \endfirsthead
 %
 \multicolumn{7}{r}{} \\ \hline
   Sequence$^*$  & Target Name  & \(l\) {$^\dagger$} & \(b\) {\(^\ddagger\)} & \(N_H\) {\(^\S\)} & date {\(^\|\)} & Exposure Time (ks) \\ \hline
        \endhead
    \hline
    \endfoot
    \hline
    \multicolumn{7}{l}{\footnotemark[$*$]	Sequence numbers of the Suzaku archive.} \\
    \multicolumn{7}{l}{\footnotemark[$\dagger$]	Galactic longitude in units of degrees.} \\
     \multicolumn{7}{l}{\footnotemark[$\ddagger$]	Galactic latitude in units of degrees.} \\
      \multicolumn{7}{l}{\footnotemark[$\S$]	Galactic hydrogen column density in units of $10^{20}{\rm cm}^{-2}$}. \\
      \multicolumn{7}{l}{\footnotemark[$\|$]	Observation start date.} \\

    %
    \endlastfoot
    100018010 & NEP & 95.74 & 28.67 & 4.0 & 2005/09/02 & 93.8 \\
    100030020 & A2218\-offset & 97.71 & 40.11 & 2.4 & 2005/10/02 & 43.1 \\
    100046010 & LOCKMANHOLE & 148.97 & 53.13 & 0.6 & 2005/11/14 & 68.1 \\
    101002010 & LOCKMANHOLE & 149.71 & 53.22 & 0.6 & 2006/05/17 & 72.9 \\
    102018010 & LOCKMANHOLE & 149.73 & 53.20 & 0.6 & 2007/05/03 & 88.3 \\
    103009010 & LOCKMANHOLE & 149.71 & 53.22 & 0.6 & 2008/05/18 & 83.4 \\
    104002010 & LOCKMANHOLE & 149.71 & 53.22 & 0.6 & 2009/06/12 & 92.8 \\
    105003010 & LOCKMANHOLE & 149.72 & 53.22 & 0.6 & 2010/06/11 & 71.3 \\
    106001010 & LOCKMANHOLE & 149.73 & 53.21 & 0.6 & 2011/05/04 & 40.3 \\
    107001010 & LOCKMANHOLE & 149.74 & 53.20 & 0.6 & 2012/05/05 & 32.2 \\
    108001010 & LOCKMANHOLE & 149.66 & 53.18 & 0.6 & 2013/11/06 & 36.0 \\
    109014010 & LOCKMANHOLE & 149.66 & 53.17 & 0.6 & 2014/11/30 & 30.2 \\
    402044010 & SW UMA & 164.79 & 36.95 & 4.1 & 2007/11/06 & 16.8 \\
    402046010 & BZ UMA & 159.03 & 38.83 & 4.8 & 2008/03/24 & 25.1 \\
    402089020 & TW HYA & 278.66 & 22.96 & 6.8 & 2007/11/25 & 19.3 \\
    403008010 & AM HERCULES BGD & 77.40 & 20.27 & 6.5 & 2008/11/01 & 40.5 \\
    403039010 & ASAS J002511+1217.2 & 112.94 & -50.08 & 5.7 & 2009/01/10 & 28.9 \\
    404035010 & HD72779 & 205.49 & 31.34 & 2.9 & 2009/11/06 & 65.2 \\
    405014010 & PSR J0614-33 & 240.49 & -21.83 & 3.9 & 2010/10/29 & 27.6 \\
    405034010 & EG AND & 121.56 & -22.18 & 13.0 & 2011/02/05 & 90.7 \\
    406007010 & 1FGL J2339.7-0531 & 81.33 & -62.46 & 3.2 & 2011/06/29 & 95.2 \\
    407039010 & EUVE J1439 +75.0 & 114.11 & 40.15 & 3.3 & 2012/05/20 & 28.3 \\
    407043010 & CH UMA & 142.92 & 42.67 & 4.7 & 2012/05/01 & 45.2 \\
    407045010 & BF ERI & 201.05 & -31.29 & 5.8 & 2013/02/27 & 30.7 \\
    408029010 & V1159 ORI & 206.54 & -19.93 & 27.6 & 2014/03/16 & 192.0 \\
    408030010 & SWIFT J2319.4+2619 & 98.49 & -32.24 & 6.8 & 2013/12/07 & 38.2 \\
    409029010 & 1RXS J032540.0-08144 & 192.85 & -48.96 & 5.9 & 2014/07/31 & 37.3 \\
    409030010 & IW ERIDANI & 216.42 & -40.63 & 2.8 & 2014/08/01 & 39.8 \\
    500026010 & NEP & 95.81 & 28.67 & 4.0 & 2006/02/10 & 80.5 \\
    500027020 & HIGH LAT. DIFFUSE B & 272.41 & -58.26 & 3.3 & 2006/02/17 & 103.6 \\
    501001010 & SKY\-50.0\--62.4 & 278.68 & -47.07 & 2.4 & 2006/03/01 & 79.1 \\
    501002010 & SKY\-53.3\--63.4 & 278.63 & -45.3 & 5.8 & 2006/03/03 & 101.5 \\
    501004010 & DRACO HVC REGION A & 91.22 & 42.39 & 1.8 & 2006/03/20 & 57.3 \\
    501005010 & DRACO HVC REGION B & 90.09 & 42.69 & 1.5 & 2006/03/22 & 57.3 \\
    501101010 & DRACO ENHANCEMENT & 93.99 & 43.97 & 1.1 & 2006/11/09 & 68.7 \\
    501104010 & MBM12 OFF-CLOUD & 157.36 & -36.82 & 9.0 & 2006/02/06 & 72.1 \\
    502047010 & LOW\-LATITUDE\-86-21 & 86.00 & -20.77 & 7.9 & 2007/05/09 & 79.6 \\
    502076010 & ERIDANUS HOLE & 213.42 & -39.10 & 2.6 & 2007/07/30 & 95.8 \\
    503104010 & ARC\-BACKGROUND & 240.49 & -66.01 & 4.1 & 2008/12/30 & 176.4 \\
    504062010 & VICINITY OF NGC 4051 & 150.12 & 70.29 & 1.2 & 2009/12/19 & 84.0 \\
    504069010 & SEP \#1 & 276.39 & -29.82 & 5.8 & 2009/11/14 & 51.9 \\
    504070010 & NEP \#1 & 96.38 & 29.78 & 4.5 & 2009/11/15 & 52.2 \\
    504071010 & SEP \#2 & 276.38 & -29.82 & 5.8 & 2009/12/05 & 58.0 \\
    504072010 & NEP \#2 & 96.40 & 29.78 & 4.5 & 2009/12/07 & 43.6 \\
    504073010 & SEP \#3 & 276.38 & -29.82 & 5.8 & 2009/12/14 & 44.4 \\
    504074010 & NEP \#3 & 96.40 & 29.78 & 4.5 & 2009/12/15 & 45.7 \\
    504075010 & SEP \#4 & 276.38 & -29.81 & 5.8 & 2009/12/27 & 49.7 \\
    504076010 & NEP \#4 & 96.42 & 29.79 & 4.5 & 2009/12/28 & 49.6 \\
    505044010 & L139\-B-32 & 138.78 & -32.31 & 6.9 & 2011/01/08 & 78.9 \\
    505058010 & L168\-B53 & 167.63 & 53.18 & 0.9 & 2010/11/19 & 77.7 \\
    506024010 & 3C 59 VICINITY 1 & 142.15 & -29.91 & 7.2 & 2012/01/14 & 48.9 \\
    506025010 & 3C 59 VICINITY 2 & 141.97 & -31.19 & 6.6 & 2012/01/26 & 156.6 \\
    506056010 & G236+38 OFF & 237.09 & 41.12 & 2.1 & 2011/06/07 & 68.0 \\
    508073010 & MBM16-OFF & 165.84 & -38.39 & 19.0 & 2013/08/09 & 77.4 \\
    509008010 & HOT BLOB 2 & 164.91 & 38.21 & 3.2 & 2015/04/25 & 72.9 \\
    509009010 & HOT BLOB 3 & 167.90 & 36.02 & 5.0 & 2015/04/27 & 75.9 \\
    700011010 & SWIFT J0746.3+2548 & 194.50 & 22.91 & 5.1 & 2005/11/04 & 88.8 \\
    701008010 & IRASF11223-1244 & 272.53 & 44.75 & 4.8 & 2006/11/25 & 33.5 \\
    701057010 & APM 08279+5255 & 165.73 & 36.24 & 4.7 & 2006/10/12 & 92.4 \\
    701057020 & APM 08279+5255 & 165.73 & 36.23 & 4.7 & 2006/11/01 & 90.9 \\
    701057030 & APM 08279+5255 & 165.78 & 36.24 & 4.7 & 2007/03/24 & 105.8 \\
    702031010 & MRK 1239 & 239.29 & 38.22 & 4.4 & 2007/05/06 & 57.9 \\
    702062010 & Q0450-1310 & 211.76 & -32.06 & 10.3 & 2008/03/10 & 15.3 \\
    702064010 & Q1017+1055 & 230.34 & 50.84 & 3.4 & 2007/11/27 & 17.5 \\
    702076010 & SWIFT J0918.5+0425 & 227.08 & 34.42 & 3.8 & 2007/11/04 & 54.4 \\
    702115010 & IRAS 10565+2448 & 212.31 & 64.73 & 1.1 & 2007/11/06 & 36.6 \\
    703002010 & PKS0208-512 & 276.08 & -61.77 & 1.9 & 2008/12/14 & 48.7 \\
    703003010 & Q0827+243 & 200.00 & 31.88 & 3.3 & 2008/10/27 & 44.8 \\
    703008010 & SWIFT J0911.2+4533 & 174.69 & 43.11 & 1.3 & 2008/10/25 & 85.9 \\
    703016010 & SWIFT J0134.1-3625 & 261.76 & -77.07 & 2.1 & 2008/05/20 & 34.9 \\
    703036020 & Q0551-3637 & 242.39 & -26.93 & 3.6 & 2008/05/14 & 21.6 \\
    703037010 & Q0109-3518 & 275.54 & -80.97 & 2.0 & 2008/05/20 & 30.0 \\
    703040010 & Q0940-1050 & 246.41 & 30.44 & 4.6 & 2008/05/30 & 29.9 \\
    703042010 & J081618.99+482328.4 & 171.04 & 33.70 & 5.8 & 2009/03/27 & 80.5 \\
    703062010 & NGC 1448 & 251.61 & -51.36 & 1.0 & 2009/02/17 & 45.2 \\
    703065010 & IRASF01475-0740 & 160.66 & -65.86 & 2.2 & 2008/07/14 & 53.8 \\
    704008010 & 1739+518 & 79.52 & 31.87 & 3.1 & 2009/06/03 & 22.5 \\
    704013010 & 2MASX J02485937+2630 & 153.12 & -29.32 & 15.2 & 2009/07/18 & 37.1 \\
    704014010 & UGC 12741 & 105.64 & -29.87 & 7.9 & 2009/06/07 & 45.7 \\
    704039010 & PKS 0326-288 & 224.92 & -55.38 & 1.0 & 2010/01/30 & 52.2 \\
    704048010 & NGC 3718 & 146.85 & 60.20 & 1.1 & 2009/10/24 & 42.9 \\
    704050010 & SDSS J1352+4239 & 88.09 & 70.11 & 1.0 & 2009/06/02 & 29.5 \\
    704052010 & SDSS J0943+5417 & 161.24 & 46.43 & 1.5 & 2009/05/24 & 32.1 \\
    704053010 & IC 2497 & 190.29 & 48.82 & 1.1 & 2009/04/18 & 74.4 \\
    705001010 & MRK 18 & 155.88 & 39.41 & 5.0 & 2010/05/14 & 35.5 \\
    705003010 & 1150+497 & 145.50 & 64.96 & 2.2 & 2010/11/12 & 99.7 \\
    705012010 & EMS1160 & 120.04 & 27.95 & 8.6 & 2010/04/26 & 20.3 \\
    705023010 & LEDA 84274 & 106.76 & 47.41 & 1.3 & 2010/05/15 & 49.5 \\
    705024010 & IRAS 01250+2832 & 132.53 & -33.41 & 8.2 & 2011/01/10 & 51.2 \\
    705027010 & EMS1341 & 102.86 & 19.43 & 21.0 & 2010/11/27 & 21.3 \\
    705045010 & IRAS 12072-0444 & 283.96 & 56.33 & 3.5 & 2010/12/04 & 56.3 \\
    705046010 & IRAS 00397-1312 & 113.94 & -75.67 & 1.8 & 2010/12/28 & 78.4 \\
    705054010 & NGC 3147 & 136.30 & 39.49 & 3.3 & 2010/05/24 & 132.5 \\
    706004010 & NGC6251\-LOBE\-BGD1 & 116.19 & 31.06 & 7.9 & 2011/04/16 & 17.5 \\
    706005010 & NGC6251\-LOBE\-BGD2 & 115.83 & 31.62 & 6.0 & 2011/04/16 & 10.9 \\
    706005020 & NGC6251\-LOBE\-BGD2 & 115.77 & 31.62 & 6.0 & 2011/08/16 & 11.2 \\
    706013010 & 3C78 & 174.83 & -44.51 & 14.6 & 2011/08/20 & 91.2 \\
    706037010 & MRK 231 & 121.78 & 60.27 & 1.0 & 2011/04/27 & 192.1 \\
    706038010 & IRAS 09104+4109 & 180.98 & 43.54 & 1.5 & 2011/11/18 & 77.4 \\
    707006010 & 3C 236 BACKGROUND & 190.38 & 53.69 & 1.0 & 2012/05/08 & 42.2 \\
    707007010 & 2FGL J0923.5+1508 & 215.99 & 40.48 & 3.2 & 2012/04/29 & 86.0 \\
    707008010 & 2FGL J1502.1+5548 & 92.72 & 52.92 & 1.4 & 2012/05/22 & 60.5 \\
    707009010 & 2FGL J0022.2-1853 & 82.11 & -79.36 & 2.1 & 2012/05/30 & 34.3 \\
    707012010 & NGC 3431 & 266.05 & 37.09 & 4.8 & 2012/06/11 & 54.7 \\
    707021010 & AO 0235+164 & 156.79 & -39.11 & 10.3 & 2013/01/18 & 39.4 \\
    707041010 & 0827+243 & 200.00 & 31.87 & 3.3 & 2012/10/13 & 7.9 \\
    708002010 & NGC 3997 & 218.79 & 77.83 & 7.1 & 2013/05/27 & 80.6 \\
    708004010 & ESO 119-G008 & 266.69 & -38.88 & 1.3 & 2013/04/29 & 97.2 \\
    708023010 & MRK533 & 90.64 & -48.80 & 5.2 & 2013/12/08 & 50.2 \\
    708026010 & NGC 235A & 94.33 & -85.93 & 1.5 & 2013/12/10 & 19.6 \\
    708038010 & IRAS F11119+3257 & 192.25 & 68.36 & 2.2 & 2013/05/13 & 241.3 \\
    708039010 & VII ZW 403 & 127.83 & 37.30 & 3.9 & 2013/12/01 & 81.8 \\
    708043010 & NGC 3660 & 269.08 & 48.37 & 4.0 & 2013/11/28 & 121.8 \\
    708044010 & B2 1023+25 & 207.08 & 57.61 & 1.7 & 2013/05/30 & 93.7 \\
    709003010 & NGC 2655 & 134.95 & 32.71 & 2.4 & 2014/05/12 & 72.2 \\
    709004010 & SWIFT J2248.8+1725 & 85.73 & -36.43 & 7.7 & 2014/12/06 & 73.0 \\
    709007010 & SWIFT J0714.2+3518 & 182.51 & 19.57 & 6.7 & 2015/04/04 & 72.9 \\
    709009010 & ARP318 & 173.93 & -64.98 & 2.8 & 2014/08/04 & 74.0 \\
    709019010 & Q0142-100 & 161.6 & -68.48 & 3.2 & 2014/07/16 & 56.7 \\
    709020010 & HE0512-3329 & 236.62 & -33.86 & 2.6 & 2014/10/02 & 6.5 \\
    709020020 & HE0512-3329 & 236.62 & -33.86 & 2.6 & 2014/10/03 & 26.5 \\
    709020030 & HE0512-3329 & 236.63 & -33.84 & 2.6 & 2015/02/18 & 24.7 \\
    709021010 & I ZW 18 & 160.55 & 44.86 & 2.7 & 2014/05/15 & 16.5 \\
    709021020 & I ZW 18 & 160.49 & 44.84 & 2.7 & 2014/10/04 & 72.9 \\
    802083010 & COMABKG & 75.61 & 83.17 & 1.0 & 2007/06/21 & 29.8 \\
    803041010 & NGC1961BACKGROUND & 145.24 & 18.80 & 13.1 & 2008/10/09 & 23.1 \\
    808057010 & BULLET-BKG & 266.17 & -20.77 & 6.8 & 2013/05/10 & 48.9 \\
    809052010 & OFF-FIELD1 & 212.28 & 55.01 & 2.1 & 2014/05/05 & 36.6 \\
    901005010 & GRB070328 & 235.21 & -44.99 & 2.9 & 2007/03/28 & 50.4 \\
    904001010 & GRB 090709A & 91.78 & 20.22 & 8.5 & 2009/07/09 & 58.4 \\
\end{longtable}


\end{document}